%% file: main-arXiv.tex
\documentclass[12pt,a4paper]{article}

\newlength\typewidth
\newlength\lfskip
\usepackage{ifthen} 
\newboolean{formpla}
\setboolean{formpla}{false} 

\newboolean{articletitles}
\setboolean{articletitles}{true} 

\usepackage{mciteplus}
\usepackage{cite}

\usepackage[breaklinks]{hyperref}    

\usepackage{caption} 

\usepackage{graphicx}

\input{preamble}

\begin{document}

\begin{titlepage}

\quad
\vspace*{3.0cm}

{\bf\boldmath\huge
\begin{center}
A Brief Guide to Exotic Hadrons
\end{center}
}

\begin{center}
Nils H\"usken,$^1$ 
Elisabetta Spadaro Norella,$^2$
and Ivan Polyakov$^3$ 
\bigskip\\
{\it\footnotesize
$ ^1$Johannes Gutenberg-Universit\"at Mainz, Mainz, Germany;\\
e-mail: nhuesken@uni-mainz.de\\
$ ^2$University of Genoa and INFN, Genoa, Italy;\\
e-mail: elisabetta.spadaro.norella@cern.ch \\
$ ^3$The University of Manchester;\\ e-mail: ivan.polyakov@cern.ch\\
}
\quad\\
\today
\end{center}

\vspace{\fill}

\begin{abstract}
\noindent
Exotic hadrons are a new class of hadronic states whose properties do not allow them to be classified as conventional quark-antiquark mesons or three quark baryons.
Finding new and understanding established exotic states is
 the most important topic in today's hadron spectroscopy and a promising avenue to advance our knowledge on Quantum Chromodynamics in the non-perturbative regime.
While several high-quality reviews on the topic exist, they are all at an advanced level.
The present article aims to address new-comers to the field with a simple introduction to exotic hadrons with an emphasis on the experimental studies. 
\end{abstract}

\vspace*{1.1cm}
\vfill

\begin{center}
{\it to appear in Modern Physics Letters A}
\end{center}
\end{titlepage}

\input{introduction}

\input{theory_short}

\input{cards_content}

\input{conclusion}


\section*{Acknowledgments}

NH acknowledges support from the European Union Horizon 2020 research and innovation programme under Marie Sk\l{}odowska-Curie grant agreement No.\ 894790, and the Helmholtz-Institute Mainz, Section SPECF.

\bibliographystyle{LHCb}
\bibliography{sample}

\end{document}

%% file: preamble.tex
\usepackage{color}
\usepackage{xspace}

\usepackage{multicol}
\usepackage{multirow}
\usepackage{hyperref}
\usepackage{comment}
\usepackage{rotating}

\usepackage{float}

\usepackage{wrapfig}

\usepackage{outlines}

\usepackage[dvipsnames]{xcolor}

\usepackage{pdflscape}
\usepackage{nicefrac}
\usepackage{enumitem}

\usepackage{amsmath}

\usepackage{afterpage}
\usepackage{lineno}

\newboolean{captionmode} 
\setboolean{captionmode}{true}

\ifthenelse{\boolean{formpla}}
{} 
{ 
\newcommand\tbl[2]{\caption{#1} \scriptsize {#2}}
\def\Hline{\hline}

\def\toprule{\Hline\Hline\\[-7.5pt]}
\def\colrule{\\[-7.5pt]\hline\\[-7.5pt]}

\def\botrule{\\[-9pt]\Hline\Hline}

}

\newcommand{\todo}[2]{ {\color{red} {[ToDo: #1] #2}} } 

\newcommand{\partcolrule}[2]{\\[-6.5pt]\cline{#1-#2}\\[-6.5pt]} 

\def\c{{\ensuremath{c}}\xspace}
\def\cbar{{\ensuremath{\overline{c}}}\xspace}
\def\b{{\ensuremath{b}}\xspace}
\def\bbar{{\ensuremath{\overline{b}}}\xspace}
\def\q{{\ensuremath{q}}\xspace}
\def\qbar{{\ensuremath{\overline{q}}}\xspace}
\def\u{{\ensuremath{u}}\xspace}
\def\ubar{{\ensuremath{\overline{u}}}\xspace}
\def\d{{\ensuremath{d}}\xspace}
\def\dbar{{\ensuremath{\overline{d}}}\xspace}
\def\s{{\ensuremath{s}}\xspace}
\def\sbar{{\ensuremath{\overline{s}}}\xspace}

\def\ee{{\ensuremath{e^+e^-}}\xspace}

\def\jpsi{\ensuremath{J/\psi}\xspace}
\def\psitwos{\ensuremath{\psi(2S)}\xspace}
\def\theX{\ensuremath{\chi_{c1}(3872)}\xspace}

\def\pip{\ensuremath{\pi^+}\xspace}
\def\pim{\ensuremath{\pi^-}\xspace}
\def\Kp{\ensuremath{K^+}\xspace}
\def\Km{\ensuremath{K^-}\xspace}
\def\proton{\ensuremath{p}\xspace}

\def\pp{\ensuremath{\proton\proton}\xspace}

\def\kaon{\ensuremath{K}\xspace}
\def\pion{\ensuremath{\pi}\xspace}

\def\Kstar{\ensuremath{K^*}\xspace}

\def\piz{\ensuremath{\pi^0}\xspace}
\def\g{\ensuremath{\gamma}\xspace}

\def\B{{\ensuremath{B}}\xspace}
\def\D{{\ensuremath{D}}\xspace}
\def\Db{\ensuremath{\overline{\D}}\xspace}

\def\Bp{\ensuremath{\B^+}\xspace}
\def\Bm{\ensuremath{\B^-}\xspace}
\def\Bz{\ensuremath{\B^0}\xspace}

\def\Bd{\Bz}

\def\Bs{\ensuremath{\B_s^0}\xspace}

\def\Lb{\ensuremath{\Lambda_b^0}\xspace}

\def\Dp{\ensuremath{\D^+}\xspace}
\def\Dm{\ensuremath{\D^-}\xspace}
\def\Dz{\ensuremath{\D^0}\xspace}
\def\Dzb{\ensuremath{\overline{\D}^0}\xspace}
\def\Ds{\ensuremath{\D_s^+}\xspace}

\def\Dstar{\ensuremath{\D^*}\xspace}
\def\Dstarz{\ensuremath{\D^{*0} }\xspace}
\def\Dstarp{\ensuremath{\D^{*+} }\xspace}
\def\Dstars{\ensuremath{\D_{s}^{*+} }\xspace}

\def\Tcc{\ensuremath{T_{\c\c}^+}\xspace}

\def\mev{\ensuremath{\mathrm{\,MeV}}\xspace}
\def\gev{\ensuremath{\mathrm{\,GeV}}\xspace}
\def\kev{\ensuremath{\mathrm{\,keV}}\xspace}
\def\tev{\ensuremath{\mathrm{\,TeV}}\xspace}

\def\fm{\ensuremath{\mathrm{\,fm}}\xspace}

\def\BelleTwo{Belle~II}

\newcommand{\est}[1]{\underline{#1}}

\def\pluszero{\nicefrac{+}{0}}

%% file: introduction.tex
\section{Introduction}

According to the quark model, hadrons are composite particles formed from quarks~\cite{GELLMANN1964214,Zweig:1964jf} which are bound together by gluons~\cite{FRITZSCH1973365}. The full theory of their internal dynamics and external interactions, Quantum Chromodynamics (QCD), is well-understood at high energies. However, at low energies, such as the mass scale of hadrons and nuclei, QCD becomes highly non-perturbative. 
Hence, describing hadrons as systems of bound quarks from first principles is a difficult task.
Therefore, progress in the field is largely driven by experimental observations. 
Over the years, the study of conventional hadrons, quark-antiquark ($q\bar q'$) mesons and three-quark ($qq'q''$) baryons has been very fruitful. Progress in both experiment and theory allowed a successful classification of these states within the constituent quark-model and lead many to believe that hadron spectroscopy is a well-understood field.
Therefore, the appearance of exotic hadrons, which are neither quark-antiquark nor three quark states, triggered large excitement. Detailed studies of their properties will provide an ultimate test for the underlying theory and deepen our understanding of the non-perturbative aspects of QCD.
Hence, the exploration of exotic hadrons is of utmost importance to our field.

Comprehensive reviews on exotic hadrons, both from an experimental and theoretical point-of-view, can be found for example in Refs.~\cite{Johnson:2024omq,Lebed:2023vnd,Huang:2023jec,PDG2023_Non_qq_review,PDG2023_Penta_review,Maiani:2022psl,Chen:2022asf,Wu:2022ftm,Gross:2022hyw,Brambilla:2019esw,Liu:2019zoy,Guo:2017jvc,Karliner:2017qhf,Olsen:2017bmm,Ali:2017jda,Lebed:2016hpi,Briceno:2015rlt,Liu:2024uxn}.
These in-depth, detailed reviews are typically addressed to experts and can be overwhelming for those looking to join a field that is constantly evolving.
With this article we aim to address beginners, giving an overview of the field and summarizing the arguments on why prominent examples of exotic hadrons are commonly viewed as such. 
This article is therefore a simple introduction into the topic with emphasis on the experimental perspective, focusing on hadrons involving \c (charm) or \b (bottom) quarks. It consists of the following sections: a short summary of the most important experimental discoveries of exotic hadrons in Section~\ref{sec:experiment}; a brief overview of theoretical models commonly used for their description in Section~\ref{sec:theory}; and an attempt to categorize all the states known to date, along with a summary on their measured properties in a kind of field guide in Section~\ref{sec:cards}.

\section{Experimental progress}
\label{sec:experiment}

\begin{table}[]
    \ifthenelse{\boolean{formpla}}
    {}{\centering}
    \tbl{Summary of current and past experiments contributing to the studies of exotic hadrons with heavy quarks.
    Future dates in the Data Taking column represent plans or intentions.
    }{
    \begin{tabular}{@{}ccccc@{}} \toprule
       Experiment  & Collisions & Energy (c.m.) & Production modes & Data Taking \\ 
    \colrule
        BaBar & \multirow{3}{*}{\ee} & \multirow{3}{*}{$10.58-11.2\gev$} 
            & \multirow{3}{*}{\ee, $\g\g$, \B-decays}
            & $1999-2008$ \\
        Belle & & & & $1999-2010$ \\
        \BelleTwo & & & & $2018-2035$ \\
    \colrule
        CDF-II & \multirow{2}{*}{$\proton\bar{\proton}$} & 
                \multirow{2}{*}{$1.96\tev$}
            & \multirow{2}{*}{$\proton\bar{\proton}$, $\b$-decays}
            & \multirow{2}{*}{$2001-2011$} \\
        D0~(Run2) & & & & \\
    \colrule
        CLEO-c & \multirow{2}{*}{\ee} 
            & $3.97-4.26\gev$
            & \multirow{2}{*}{\ee, $\g\g$} & $2003-2008$ \\
        BESIII & & $1.8-4.96\gev$ & & $\phantom{+}2008-2030+$  \\
    \colrule
        ATLAS & \multirow{3}{*}{$\proton\proton$ (+ $\proton$Pb, PbPb)} & 
                \multirow{3}{*}{$7-13.6\tev$}
            & \multirow{3}{*}{$\proton\proton$, $\b$-decays}
            & \multirow{3}{*}{$2010-2041$} \\
        CMS & & & & \\
        LHCb & & & &  \\
    \colrule
        COMPASS & \multirow{2}{*}{$\mu\proton$, $\pi\proton$} & 
        \multirow{2}{*}{$p_{beam}=160-200\gev/c$}
        & \multirow{2}{*}{diffractive} & $2002-2021$ \\
        AMBER & & & & $\phantom{+}2023-2032+$ \\
        GlueX & $\g N$& 
        $E_\gamma=8.0-11.5\gev$
        & photoproduction & $\phantom{+}2015-2028+$ \\
    \botrule
    \label{tab:experiments}
    \end{tabular}
    }
\end{table}

Particles with a quark content that differs from $q\bar q'$ mesons and $qq'q''$ baryons have been widely discussed since the birth of the constituent quark model in 1964~\cite{GELLMANN1964214,Zweig:1964jf}. 
For a long time such states, referred to as exotic hadrons, remained merely a hypothesis.
In the 1990's several scalar tetraquark candidates emerged in the light quark (\u, \d and \s) sector, but the large natural widths of many of these states in a region densely populated with conventional mesons made an unambiguous theoretical interpretation difficult~\cite{PDG2021_ScalarMesonsBelow1GeV}. 
Compelling evidence for exotic hadrons emerged only years later with the inclusion of heavy quarks (\c and \b) into the studies.

\begin{figure}[!htb!]
    \centering
    \includegraphics[width=1.0\textwidth]{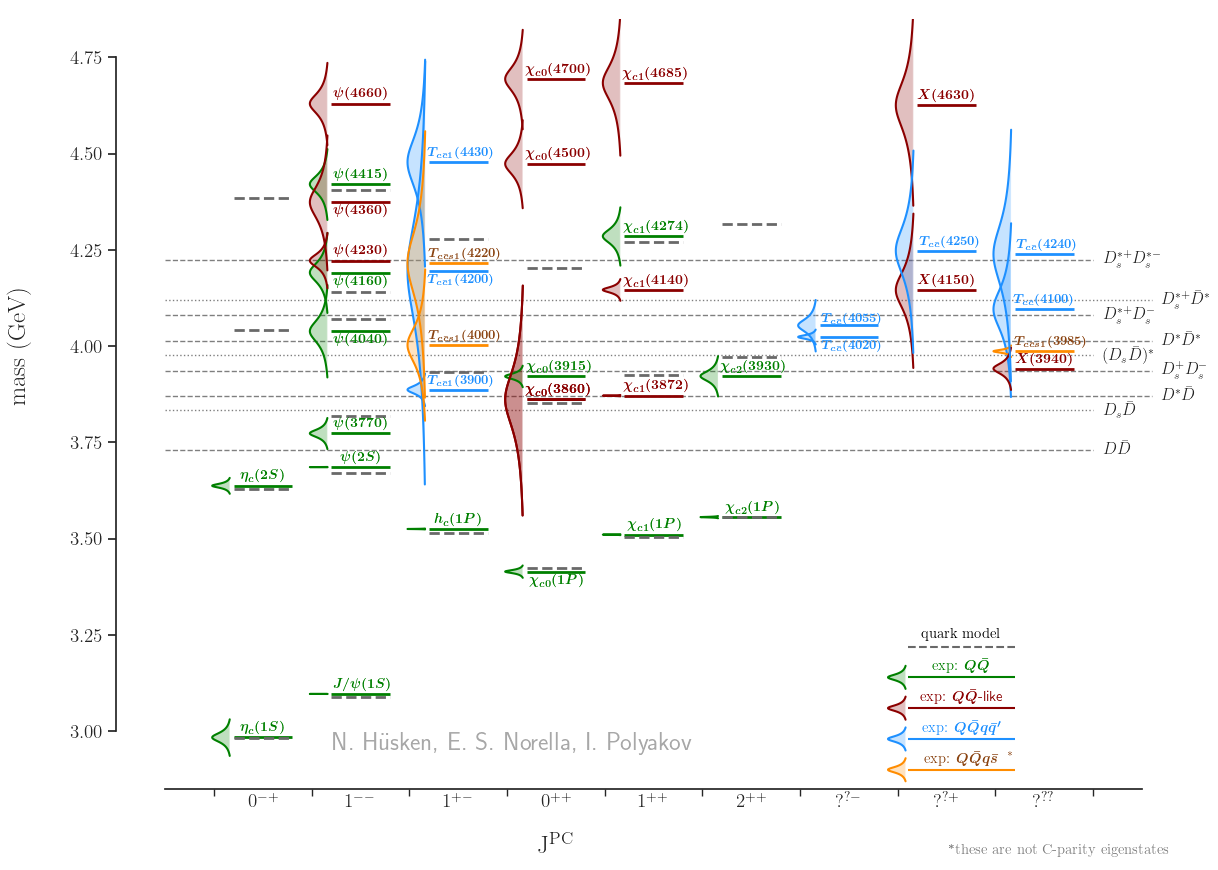}
    \caption{Spectrum of conventional and potentially exotic charmonium(-like) states. Potential model calculations for conventional charmonium are overlaid with experimental measurements.}
    \label{fig:specQQ}
\end{figure}

In 2003 the first tetraquark candidate \theX was discovered by the Belle collaboration~\cite{PhysRevLett.91.262001}.
Despite the discovery being in the decay to $\jpsi\pip\pim$, clearly signaling a constituent $\c\cbar$ pair, a conventional charmonium interpretation is considered highly unlikely.
One important part in the exotic interpretation of the \theX is the higher level of theoretical understanding of the conventional charmonium ($\c\cbar$) spectrum (see Fig.~\ref{fig:specQQ}) in comparison to the light quark sector. 
Another key argument is the small width of the \theX of the order of an \mev, much smaller than expected for a conventional charmonium state above the open-charm threshold.
This discovery was quickly confirmed by other experiments -- BaBar~\cite{BaBar:2004oro} at SLAC, as well as CDF~\cite{CDF:2003cab} and D0~\cite{D0:2004zmu} at Tevatron -- and sparked the renaissance of hadron spectroscopy.
Soon, more exotic signals with a hidden $\c\cbar$ pair, including manifestly exotic states with minimal quark content $\c\cbar\u\dbar$~\cite{Belle:2007hrb}, were discovered, accompanied by the analogous structures with hidden $\b\bbar$ pairs~\cite{Belle:2011aa}.
At first, the progress in the field was lead by the experiments at B-factories~\cite{BaBar:2014omp} -- Belle, BaBar and CLEO, being complemented by the results from the CDF and D0 experiments at Tevatron~\cite{Bandurin:2014bhr}.
In 2008 the BESIII experiment~\cite{BESIII:2020nme} started collecting data in \ee collisions at center of mass energies of up to 4.6\gev (up to 5 GeV nowadays) and provided a unique probe for a large list of exotic hadrons with hidden charm.
When the LHC began operation in 2010, the LHCb, CMS and ATLAS experiments joined the exploration of the exotic hadrons, of which it was LHCb who
has established itself as the world’s leading experiment in the field.
Among its many results in hadron spectroscopy,
a few of the most outstanding are the discoveries of the first pentaquarks with hidden charm~\cite{LHCb:2015yax}, the first tetraquarks with single~\cite{LHCb:2020bls,LHCb:2020pxc} and double open charm~\cite{LHCb-PAPER-2021-031}, and the first tetraquarks formed exclusively by heavy quarks~\cite{LHCb:2020bwg}.
Many new exotic hadron candidates observed by LHCb will require next-generation experiments for independent confirmation.
In 2018 the \BelleTwo \ experiment~\cite{Belle-II:2018jsg} started collecting data and has already produced first results on exotic hadrons, promising much more in the future.
Brief information on the current and past experiments contributing to the field is presented in Table~\ref{tab:experiments}.

Important progress has also been achieved in the light-quark sector by experiments like COMPASS~\cite{COMPASS:2007rjf}, GlueX~\cite{GlueX:2020idb} and BESIII. However, the interpretation of experimental measurements in that region is in general more difficult 
as the states are broad and overlap, rendering many observations model-dependent.
Therefore, in this review, we will focus on exotic hadrons containing at least one heavy quark.

Various experimental techniques allow to extend the reach and perform verifications of individual measurements.
For example, the \theX state has been observed both in decays of \b-hadrons, in \ee, in hadron ($pp$, $p\overline{p}$) and even in heavy-ion collisions 
-- we can hence be fully confident it is a genuine hadronic state, and use its production properties~\footnote{like cross-section, dependence on transverse momenta and multiplicity in the event.} in the different processes to learn about its nature.
However, quite a few of the states with hidden charm are only observed in \b-hadron decays. 
They are usually identified as either narrow (${\sim}10\mev$) peaks in one-dimensional distributions or as structures (with width up to ${\sim}300\mev$) in complex multi-dimensional amplitude analyses. 
In turn, some of the states with hidden charm and all states with hidden bottom flavour are only observed in \ee annihilation.
Doubly-charmed and fully-heavy tetraquark candidates are so far observed only in prompt production in $pp$ collisions, although it should be noted that 
it is currently simply not possible to produce them in other processes with large rates.
We emphasize the need for independent confirmation of newly observed exotic hadron candidates in different production processes and decay modes. In many cases it will require new experiments like the Super-Tau-Charm Facility~\cite{Achasov:2023gey}, PANDA~\cite{Peters:2017kop} or potential experiments at future Higgs/Z-boson factories~\cite{CEPCStudyGroup:2018ghi,FCC:2018byv}.

Today, there are around 50 exotic hadron candidates discovered in experiments around the world, offering a diverse range of structures that is summarized figuratively in Fig.~\ref{fig:massif}. In the following (Section~\ref{sec:cards}), we will provide an almost zoological field guide to the exotic hadrons. Before, it is necessary to briefly summarize the different theoretical models that aim to classify (a subset of) the observed phenomena.

\afterpage{
\begin{landscape}
\begin{figure}
    \includegraphics[width=1.00\linewidth]{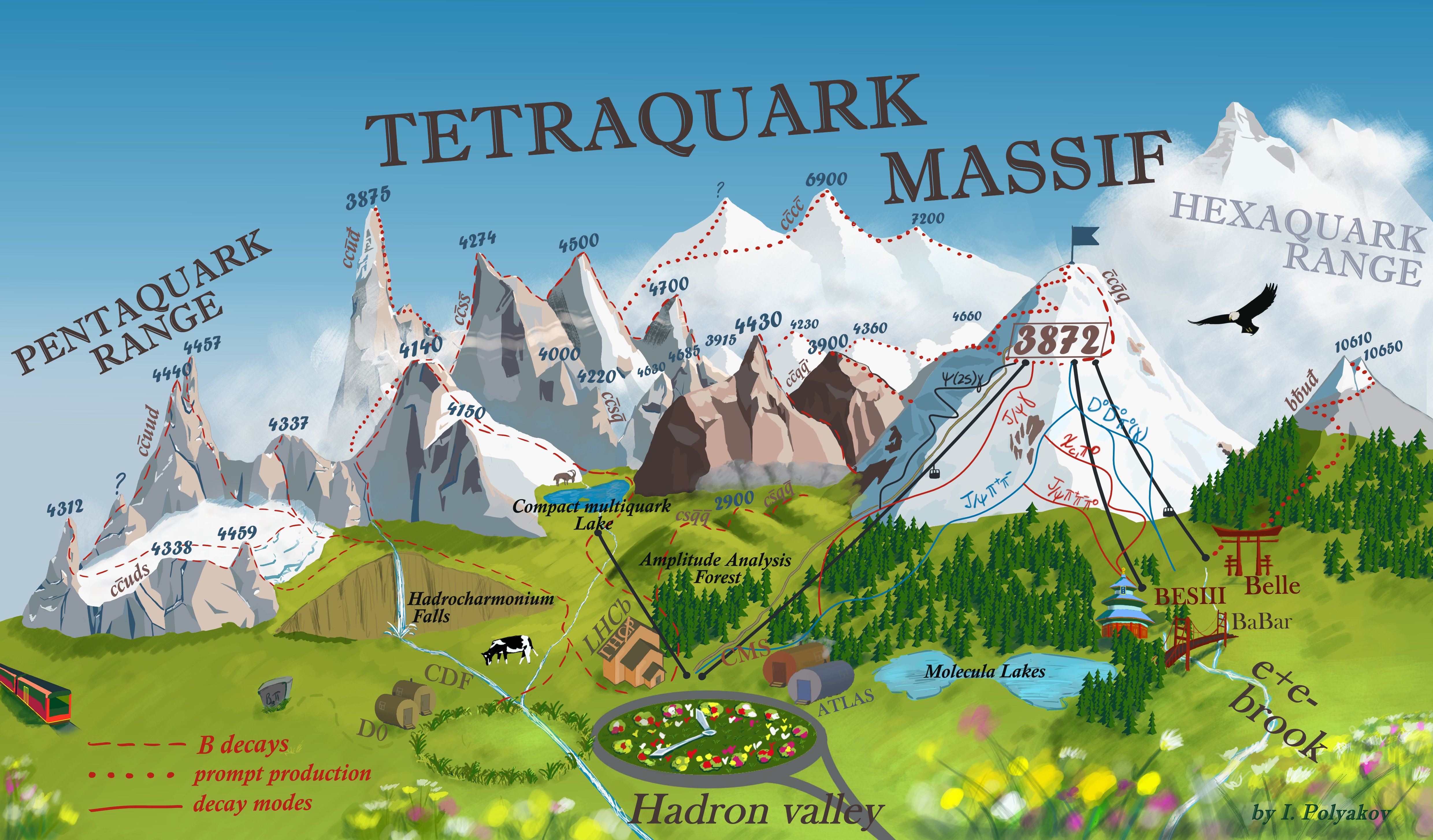}
    \caption{Illustration of exotic hadron experimental studies.}
    \label{fig:massif}
\end{figure}
\end{landscape}
}

%% file: theory_short.tex
\section{Theory perspective}
\label{sec:theory}

In this chapter, we aim to provide a concise introduction to the main theoretical concepts in the description of exotic hadrons -- for more in-depth and comprehensive overviews we refer the interested reader to articles in Refs.\cite{Lebed:2023vnd,Huang:2023jec,PDG2023_Non_qq_review,PDG2023_Penta_review,Maiani:2022psl,Chen:2022asf,Wu:2022ftm,Gross:2022hyw,Brambilla:2019esw,Liu:2019zoy,Guo:2017jvc,Karliner:2017qhf,Olsen:2017bmm,Ali:2017jda,Lebed:2016hpi,Briceno:2015rlt,Liu:2024uxn, Berwein:2024ztx}. 

Due to the highly non-perturbative nature of QCD at hadronic mass scales, understanding the internal structure of (exotic) hadrons and the interactions between them from first principles, i.e. from the Lagrangian, is an extremely challenging task.
Therefore, to provide a description of multi-quark systems one has to either rely on phenomenological approaches or the discretized version of QCD, Lattice-QCD~\cite{LatticeQCD_HashimotoOnogi_2004, LatticeQCD_Prelovsek_2017,PDG2021_LatticeQCD}.
The major progress in phenomenological approaches is related to hadrons with heavy quarks (\c or \b), because their presence limits non-perturbative contributions and relativistic effects hence simplifying the description of the system. 
Theory calculations of the full spectrum of the charmonium~\cite{Barnes:2005pb} and bottomonium~\cite{Godfrey:2015dia} mesons (systems of $\c\cbar$ and $\b\bbar$ quarks),
which rely on a phenomenological quark-quark interaction
potential~\footnote{variations of the Cornell potential~\cite{Eichten:1978tg,Eichten:1979ms}}
are readily available, and, for charmonium, are presented in Fig.~\ref{fig:specQQ}.
Given the excellent agreement with experiment, a
conventional interpretation of many exotic hadron candidates can be disfavored if no nearby state is predicted.

Lattice-QCD has emerged already in 1974~\cite{Wilson:1974sk} and with the advent of modern computers has proven itself as a particularly powerful tool to access first-principle calculations in the non-perturbative regime.
In this framework, QCD is formulated on a finite space-time grid (lattice) with fermions existing in the nodes and gluon fields in the links. Physical quantities are extracted from numerical calculations of the action integral and derived correlations.
Such methods, however, have difficulties in describing systems with both light and heavy quarks due to the need to use a large lattice with a small cell size~\cite{LatticeQCD_HISQ_2007, LatticeQCD_Jpsi_2012}.
Nevertheless, lattice results for the charmonium~\cite{HadronSpectrum:2012gic,DeTar:2018uko,Padmanath:2015era,Prelovsek:2020eiw,Piemonte:2019cbi,Cheung:2016bym} and bottomonium~\cite{Ryan:2020iog} spectrum, in part including predictions for exotic multi-quark or hybrid states, are available and largely compare very well with experiment and potential model calculations. In the future, it is to be expected that Lattice-QCD will play a leading role.

For now, Lattice-QCD and phenomenological approaches can be viewed as highly complementary in the attempt to understand the experimental spectrum of (exotic) hadrons and their properties. The two most common concepts aiming to explain the observations
are so-called {\it compact~\footnote{often also called genuine or tightly-bound} multiquark} and {\it molecular} states.

In models of {\it compact multiquark} states, quarks in an (exotic) hadron are tightly-bound
through direct interaction with the color charge of every other quark. 
The overall interaction potential in the system can be assumed to be a sum of individual quark-quark interaction terms. It can then be attempted to solve the corresponding Schr\"{o}dinger equation~\cite{Semay:1994ht}.
The system can be simplified by considering that pairs of quarks, if in a color anti-triplet state~\footnote{as an example consider two quarks with red and green color charge yielding an effective color charge anti-blue; in the language of $SU(3)_{color}$, $\mathbf{3} \otimes \mathbf{3} \to \mathbf{\bar{3}} \oplus \mathbf{6}$, with the $\mathbf{\bar{3}}$ configuration corresponding to an attraction and $\mathbf{6}$ to a repulsion between two quarks}, 
are attracted to each other.
Thus, in a multiquark object, pairs of quarks will form diquarks which then interact with the remaining constituents in the same way an anti-quark would. 
Tetra- and pentaquark states can then be viewed as $[\q_1\q_2][\qbar_3\qbar_4]$ and $[\q_1\q_2][\q_3\q_4]\qbar_5$ configurations, such that their color-interaction is effectively the same as for mesons and anti-baryons, respectively.
The resulting multiquark hadrons are expected to have similar sizes to conventional mesons and baryons, i.e. of order $1\fm$, thus the term {\it compact}.
These concepts were first discussed in Ref.~\cite{Lichtenberg:1967zz} in the context of baryons, and later in the discussion of the light scalar mesons in Ref.~\cite{Jaffe:1976ig}. A first application to exotic hadrons in the quarkonium sector was described in Ref.~\cite{Maiani:2004vq,Maiani:2014aja}, with modern dynamical diquark calculations offering explanations for exotic mesons and baryons alike~\cite{Brodsky:2014xia,Lebed:2015tna}. 
For more detailed information we refer to a recent review on the topic~\cite{Huang:2023jec} and references therein.

In the {\it molecular} picture, multiquark exotic states are described as
systems of two color-neutral objects -- in the simplest case, meson-meson or meson-baryon -- which are interacting via a QCD analogue of the van-der-Waals force, an idea that is natural in the light of the deuteron as a bound state of a proton and a neutron.
Such states are expected to have binding energies of order \mev and be large compared to conventional mesons and baryons, strongly resembling the deuteron.
In effective theories, the binding is usually explained via the exchange of a light meson like the pion\footnote{with the interaction strength and range related to the mass of the exchange particle}, 
with unknown short-range interactions accounted for by effective contact terms~\cite{Baru:2015nea}.
The idea of hadronic molecules including charm-quarks first appeared in the 1970's~\cite{Voloshin:1976ap,DeRujula:1976zlg}, predating experimental indications by three decades. 
It was revived with the observation of multiple charmonium-like exotic states whose masses lie close to two-body thresholds.
Prominent examples are the $\chi_{c1}(3872)$, whose mass is at current precision indistinguishable from the ${D}^0\bar{D}^{\ast0}$ threshold, or the recently discovered \Tcc with a mass only a fraction of an \mev below the $\Dz\Dstarp$ threshold.
The fact that these states are dominantly decaying via the $\D\Dstar$ channel further advocates for at least a sizable molecular component.
For a more detailed discussion of molecular states, we refer to recent reviews Ref.~\cite{Guo:2017jvc,Wu:2022ftm} and references therein.

Both {\it compact multiquark} and, in part, {\it molecular} models predict a large number of states in addition to those already observed.
In an ideal world, calculations in the {\it compact multiquark} and {\it molecular} pictures would not only describe a number of existing exotic hadron candidates, but provide concrete predictions of production strengths, decay patterns, or existence of partner states that can be tested in experiments to discern the different interpretations. At present, while certainly one interpretation might be preferred over the other for any given exotic hadron, for none of the existing candidates the nature has been unambiguously determined.
It is becoming more clear that even in the {\it compact multiquark} picture, nearby two-body thresholds can have a sizeable influence on a state, and thus need to be taken into account in a model.
Of course it is entirely possible, even quite likely, that most of the exotic hadrons are mixtures of {\it compact} and {\it molecular} configurations\footnote{possibly with an admixture of conventional states}, such that determining the relative fractions becomes one of main tasks in hadron spectroscopy. 

In the literature, a number of further explanations for exotic hadron candidates can be found.
In the {\it hadroquarkonium} picture, a multiquark state is described as a color-singlet $\c\cbar(\b\bbar)$ core in a specific state (spin, radial and orbital excitation) surrounded by a light-quark cloud. It is argued that decays that keep the heavy-quark core intact would be strongly favored in this scenario, which would naturally explain why some exotic hadrons seemed to preferentially decay to specific charmonium final-states~\cite{Dubynskiy:2008mq}. Prime examples are the $\psi(4230)$ and $\psi(4360)$ decaying to $J/\psi \pi\pi$ and $\psi(2S)\pi\pi$, if they were to be interpreted as hadrocharmonia with a spin-1 core in $n=1$ and $n=2$ radial excitations. 
However, the observation of decays also to final states with spin-0 $c\bar c$-cores ($\eta_c(1S)$ and $h_c(1P)$) have rendered this class of models less interesting.

Hadrons with a gluon field in an excited state, namely {\it hybrid mesons} $[\q\q g]$ and {\it glueballs} $[gg],~[ggg]$, have been hypothesized since the early days of the quark-model~\cite{Jaffe:1975fd}. 
Candidates for hybrids exist in both light-quark ($\pi_1(1600)$, $\pi_1(2015)$~\cite{Meyer:2015eta})
and quarkonium ($\psi(4230)$, $\psi(4360)$, $\Upsilon(10753)$, $\Upsilon(11020)$~\cite{TarrusCastella:2021pld, Berwein:2024ztx}) sectors.
Similar to hadroquarkonium, the concept of heavy-quark spin-symmetry also applies to decays of quarkonium-hybrids, such that one would expect the heavy-quark core to remain intact when the gluon hadronizes.
While predictions for the higher excitations of glueball states can reach into the charmonium mass range, we would expect a $[gg]$ or $[ggg]$ state at such masses to be very broad and thus refrain from further discussion here.

In some specific cases, the resonant nature of the observed hadron candidates is itself questioned. 
In particular, the presence of nearby two-body thresholds can create structures, so-called {\it cusps}, which might be mistaken for a resonant peak. 
These effects can be further enhanced in the case of a contribution from a logarithmically divergent three-point loop diagram, which produces so-called {\it triangle singularities}. 
Such kinematic effects have been proposed to explain multiple new hadron candidates, including those with heavy quarks\footnote{$\theX$, $T_{\c\cbar}$ and $T_{\b\bbar}$ tetraquark candidates near the $D^\ast \bar{D}^{(\ast)}$ and $B^\ast\bar{B}^{(\ast)}$ thresholds or the $P_{\c\cbar}$ pentaquark candidates near the $\Sigma_cD^{(\ast)}$ thresholds}.
A recent review can be found in Ref.~\cite{Guo:2019twa}.
One possibility to disentangle resonant interpretations from kinematic effects like triangle singularities is 
the observation in different production processes with entirely different kinematics. This stresses the importance of such tasks as the search for pentaquarks
in $\gamma p\to J/\psi p$ at Jefferson Lab~\cite{GlueX:2019mkq,Strakovsky:2023kqu,JointPhysicsAnalysisCenter:2023qgg,GlueX:2023pev,Duran:2022xag}.

%% file: cards_content.tex
\section{Field guide}
\label{sec:cards}

To date around 50 exotic hadrons with at least one heavy (\c or \b) quark are reported, as summarized in Table~\ref{tab:all} and represented in Fig.~\ref{fig:massif}.
Some of these states are well established while some are only candidates whose existence and resonant nature are yet to be confirmed.
Given the number of states on this list, it deserves to be called a ``new particle zoo'' in analogy to how the wide variety of conventional mesons and baryons was previously characterized.
%
Following this zoo metaphor, the experimental exploration of exotic hadrons can thus be compared to the assembly of a zoological field guide. Identifying different species and patterns will then
%
allow us to gain a deeper understanding of the underlying physics, namely QCD in the highly non-perturbative regime.
In the following, our interpretation for such an emerging field guide is presented.
Information about states (or groups of states where we deem that adequate) is presented in the form of short information cards followed by a description for in-depth reading, according to the grouping and order presented in Table~\ref{tab:all}.
We group in categories by the following criteria: number of quarks, flavor content and isospin ($I$), followed by additional properties like spin-parity, 
production and decay channels.
While aiming to provide a complete picture of the experimental knowledge of the field, we favor brevity and simplicity of description and therefore may omit details wherever appropriate.
For a full list of all reported exotic hadron candidates and their measured properties, we refer to the Review of Particle Physics \cite{Workman:2022ynf}, which we will refer to just as the ``PDG''.

\subsection{Remarks}

Below, several remarks on the notation used throughout the article are provided.

We adhere to the 2023 edition of the PDG {\bf naming scheme} for hadrons.
Where applicable, alternative names, by which these particles were previously known, and/or full names in LHCb convention~\cite{Gershon:2022xnn} are also provided for easier comparison with previous works.
Charge conjugation for particle names and mass thresholds is implied.

{\bf Underscore} is used to mark states we consider well established -- those states seen by either more than one experiment or in more than one production or decay mode. In many cases, this corresponds to a dot in the PDG Listings. If conflicting reports exist from different experiments, we refrain from using this label.

{\bf Minimal quark content} simply lists constituent quarks, but is not an attempt to describe a wave-function in flavour-space.

{\bf Experiments} in which states are observed are presented in chronological order, meaning the 
experiment that discovered the state, or the first state in a group, is listed first.

For a group of several states, the lists of {\bf decay modes} and {\bf production} channels 
is a combined list of all states from the group. In other words, it is not meant that every state is seen in every decay mode or production channel. Details may be given in the individual descriptions.

{\bf Nearby thresholds} are indicated based on closeness of masses and appropriate quark content and do not necessarily have any relevance to the particular states.

For {\bf characteristic widths} of a group of several states, an interval from minimum to maximum of the measured widths is provided. Only central values of the measurements are used.

The {$I$, $G$ and $C$} quantum numbers indicated for charged states $T_{c\bar{c}}^+$ or $T_{b\bar{b}}^+$ correspond to those of their neutral partners. In case the neutral partners are not yet observed $I$, $G$ and $C$ should be considered as requiring confirmation.

The plots presented alongside the information on the (group of) states do not necessarily represent the discovery of a given state, but instead show a few examples intended to give a general idea on the experimental observation.

\input{card_preamble}

\begin{table}[!htb!]
\ifthenelse{\boolean{formpla}}
{}{\centering}
\tbl{All known exotic hadron candidates up to date. 
States we consider well-established are underscored.
}
{
\begin{tabular}{@{}cccc@{}} \toprule
\multicolumn{3}{c}{Category} & States / Candidates \\
\colrule
\multirow{21}{*}{Meson-like} & \multirow{13}{*}{Hidden Charm} & \multirow{7}{*}{$I=0$} & $\chi-$like: \est{\theX}, \\

\multirow{21}{*}{(incl. tetraquarks)} &  & & $\chi_{c0}(3860)$, \est{$\chi_{c0}(3915)$}, \est{$\chi_{c2}(3930)$}, $X(3940)$
  \\
 
     \partcolrule{4}{4}
 &  & & $\psi-$like: \est{$\psi(4230)$}, \est{$\psi(4360)$}, $\psi(4660)$ \\

     \partcolrule{4}{4}
 &  & & with $\s\sbar$: \est{$\chi_{c1}(4140)$}, \est{$\chi_{c1}(4274)$}, \\
  &  & & $\chi_{c1}(4685)$, $\est{\chi_{c1}(4500)}$, $\chi_{c1}(4700)$ \\
  &  & &  $X(4150)$, $X(4630)$, $X(4740)$ \\

     \partcolrule{3}{4}
 &  & \multirow{5}{*}{$I=1$} & seen in \ee: \est{$T_{\c\cbar1}(3900)^{+/0}$}, \\
 &  &  &  \est{$T_{\c\cbar}(4020)^{+/0}$}, $T_{\c\cbar}(4055)^+$  \\
     \partcolrule{4}{4}
 &  &  & seen in \B decays: $T_{\c\cbar}(4050)^+$, $T_{\c\cbar}(4100)^+$,  \\
 &  &  & $T_{\c\cbar1}(4200)^+$, $T_{\c\cbar}(4240)^+$, $T_{\c\cbar}(4250)^+$, \est{$T_{\c\cbar1}(4430)^+$} \\

     \partcolrule{3}{4}
 &  & $I=\nicefrac{1}{2}$ & $T_{\c\cbar\s}(3985)^-$, $T_{\c\cbar\s1}(4000)^{-/0}$, $T_{\c\cbar\s1}(4220)^-$ \\

     \partcolrule{2}{4}
 & \multirow{3}{*}{Hidden Bottom} & $I=0$ & $\est{\Upsilon(10753)}$, $\est{\Upsilon(10860)}$, $\est{\Upsilon(11020)}$ \\
     \partcolrule{3}{4}
 & & $I=1$ & \est{$T_{\b\bbar1}(10610)^+$}, \est{$T_{\b\bbar1}(10650)^+$} \\

     \partcolrule{2}{4}
 & Hidden Double Charm &  & $T_{\c\cbar\c\cbar}(6550)$, \est{$T_{\c\cbar\c\cbar}(6900)$}, $T_{\c\cbar\c\cbar}(7290)$ \\

     \partcolrule{2}{4}
 & \multirow{3}{*}{Open Single Charm} &  & $\D_{s}^*$-like: \est{$\D_{s0}^*(2317)^+$}, \est{$\D_{s1}(2460)^+$}\\

     \partcolrule{3}{4}
 &  &  & $T_{\c\s/\c\sbar}$: $T_{\c\s0}(2900)^0$, \\
 &  &  &  $T_{\c\sbar0}(2900)^{0/++}$, $T_{\c\s1}(2900)^0$ \\

     \partcolrule{2}{4}
 & Open Double Charm &  & $T_{\c\c}(3875)^+$ \\

\colrule
     
 \multirow{3}{*}{Baryon-like} & \multirow{4}{*}{Hidden Charm} & \multirow{2}{*}{$I=\nicefrac{1}{2}(\nicefrac{3}{2})$} & $P_{\c\cbar}(4312)^+$, $P_{\c\cbar}(4440)^+$, $P_{\c\cbar}(4457)^+$ \\
 \multirow{3}{*}{(incl. pentaquarks)} &  & & $P_{\c\cbar}(4380)^+$, $P_{\c\cbar}(4337)^+$ \\
 
     \partcolrule{3}{4}
  & & $I=0(1)$ & $P_{\c\cbar\s}(4458)^0$, $P_{\c\cbar\s}(4338)^0$ \\
\botrule
\end{tabular}
}
\label{tab:all} 
\end{table}


\clearpage 

\input{card_X3872} 
\input{card_chic2p}
\input{card_psi4230}
\input{card_JpsiPhi}

\input{card_Zc3900}
\input{card_Zc_fromB}

\input{card_Zcs}

\input{card_upsilon_3states}
\input{card_Zb}

\input{card_JpsiJpsi}

\input{card_Dstar}
\input{card_Tcs}

\input{card_Tcc}


\input{card_Pc}

\input{card_Pcs}

%% file: card_preamble.tex

\def\aka{also known as }

\def\cardlength{1.0\textwidth}

\def\cardstart{
}

\def\cardend{
\vfill
\null
\vspace{-\baselineskip}
}

\newcommand{\cardtitle}[2]{
\subsection{#1 }
\vspace{-1mm}
{\small \sc \color{Periwinkle} \MakeLowercase{#2}}
}

\newcommand{\carddescription}[1]{
{\it \small \color{CadetBlue} \noindent #1}
\vspace{1mm}
}

\newcommand{\carddescriptiontop}[1]{
\ifthenelse{\boolean{formpla}}
{ 
{\it \small \color{CadetBlue} \begin{flushright}  #1 \phantom{0} \end{flushright}}
\vspace{-14pt}
}{ 
{\it \small \color{CadetBlue} \begin{flushright} #1 \phantom{0} \end{flushright}}
\vspace{-1.5\baselineskip}
}
}

\newcommand{\cardproperty}[2]{
{\bf \noindent #1:}
\setlength\hangindent{1.5em}
\hspace{-0.2em}#2
}

\ifthenelse{\boolean{formpla}}
{ 
\newcommand{\cardnamelist}[1]{
\begin{itemize}[leftmargin=7mm]
\vspace{-0.7em}
    #1    
\end{itemize}
\vspace{-0.5em}
}
}{ 
\newcommand{\cardnamelist}[1]{
\begin{itemize}[leftmargin=7mm]
\vspace{-0.6em}
\setlength\itemsep{-0.3em}
    #1    
\end{itemize}
\vspace{-0.6em}
}
}

\newcommand\cardnamealternative[1]{
{\it \small \color{Black!70} \hspace{1em} #1} 
\vspace{0.3em}
}

\newcommand\cardnotes[1]{
\ifx\reallybrief 1
\else
\vspace{5mm}
#1
\fi
}

\newcommand\scslash{\stretchrel*{$/$}{\textsc{e}}}

\ifthenelse{\boolean{formpla}}
{ 
\def\midplotoffsettop{\vspace{-2mm}}
\def\midplotoffsetbottom{\vspace{-7mm}}
\def\midplotoffsetcaption{\vspace{-2mm}}
\def\wrapcaptionoffset{\vspace{-7mm}}
\def\midplotheight{30mm}
}{ 
\def\midplotoffsettop{\vspace{0mm}}
\def\midplotoffsetbottom{\vspace{-5mm}}
\def\midplotoffsetcaption{\vspace{0mm}}
\def\wrapcaptionoffset{\vspace{-7mm}}
\def\midplotheight{35mm}
}


%% file: card_X3872.tex
\vtop{

\cardstart

\cardtitle{The $\chi_{c1}(3872)$ 
\cardnamealternative{(\aka $X(3872)$)} }{Meson-like/Hidden Charm/Isoscalar}

\begin{wrapfigure}[0]{r}{0.27\textwidth}
    \vspace{-10mm} 
    \ifthenelse{\boolean{captionmode}}{}{
    {\color{black} \footnotesize \phantom{00} Discovery by Belle~\cite{Belle:2003nnu}} \\}
    \includegraphics[width=0.27\textwidth]{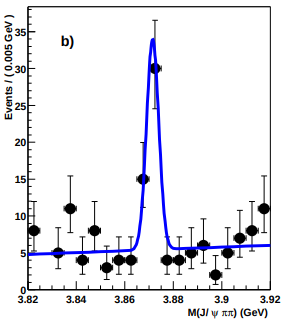}
\ifthenelse{\boolean{captionmode}}{
\wrapcaptionoffset
\caption{Discovery of the \theX by Belle~\cite{Belle:2003nnu}.}
}{}
\end{wrapfigure}

\cardproperty{quantum numbers}{$I^G(J^{PC}) = 0^+(1^{++})$}

\cardproperty{minimal quark content}{$[c\bar{c}]$, more likely $[c\bar{c}(u\bar{u}+d\bar{d})]$}

\cardproperty{experiments}{
Belle, CDF, D0, BaBar, LHCb, CMS,
\\
ATLAS, BESIII
~~(and potentially E705, COMPASS
)
}

\cardproperty{production}{\Bp,
\Bd, \Bs and \Lb decays, \\
prompt \pp, $\proton\bar{\proton}$,
$\textrm{pPb}$ ($\textrm{Pbp}$) 
and $\textrm{Pb}\textrm{Pb}$ collisions,\\ 
\noindent $\ee\to \gamma \chi_{c1}(3872),\, \omega \chi_{c1}(3872)$  \\
with the first likely via $\psi(4230)$
}

\cardproperty{decay modes}{$\pi^+\pi^- J/\psi$, $\omega J/\psi$, $D^{\ast 0}\bar{D}^0$, $\pi^0 \chi_{c1}(1P)$, \\
$\gamma J/\psi$, $\gamma \psi(2S)$}

\cardproperty{nearby threshold}{$D^{\ast 0}\bar{D}^0$}

\cardproperty{width}{$1.19\pm 0.21$ MeV
(in $\pip\pim\jpsi$ channel)
}

}

\ifthenelse{\boolean{formpla}}{
\def\pictW{125}
\def\pictH{30}
}{
\def\pictW{150}
\def\pictH{38}
}

\begin{figure}[H]
\centering
\setlength{\unitlength}{1mm}
\midplotoffsettop
\begin{picture}(\pictW,\pictH)
\ifthenelse{\boolean{captionmode}}{\vspace{-1mm}}{}
\ifthenelse{\boolean{formpla}}{
\put(0,0){\includegraphics*[height=30mm]{"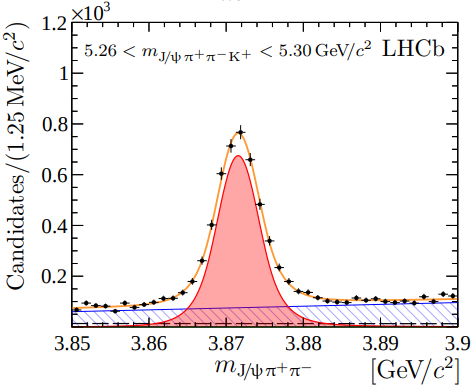"}}
\put(42,-3){\includegraphics*[height=32mm]{"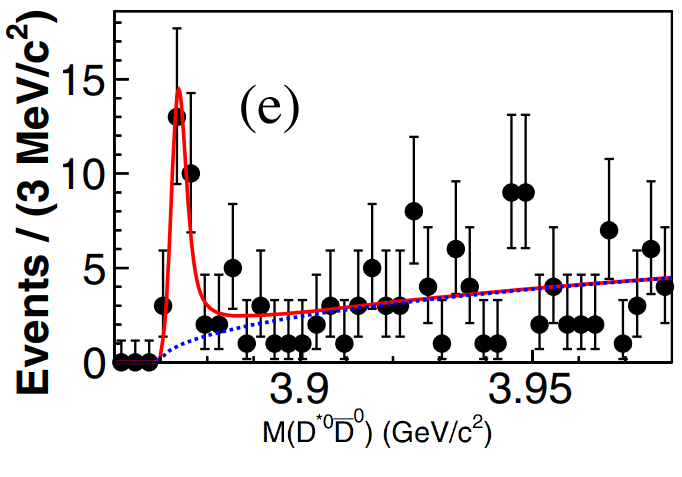"}}
\put(90,-1){\includegraphics*[height=30mm]{"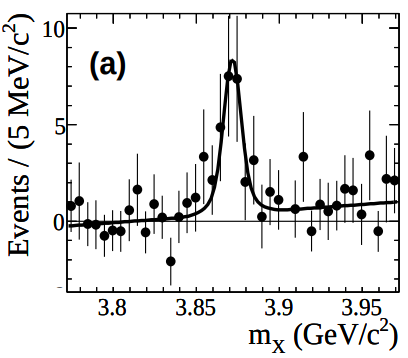"}}
}{
\put(0,0){\includegraphics*[height=38mm]{"./plots/X3872_Jpsipipi_LHCb.png"}}
\put(50,-3){\includegraphics*[height=40mm]{"./plots/X3872_DDpi0_BESIII_bottom.png"}}
\put(110,-1){\includegraphics*[height=38mm]{"./plots/X3872_Jpsig_BaBar_crop.png"}}
}
\ifthenelse{\boolean{captionmode}}{}{
\put(7,32){\footnotesize in $\jpsi\pip\pim$ at LHCb~\cite{LHCb:2020fvo}}
\put(47,32){\footnotesize in $\Dstarz[\to\Dz\piz]\Dzb$ at BESIII~\cite{BESIII:2020nbj}}
\put(97,32){\footnotesize in $\jpsi\g$ at BaBar~\cite{BaBar:2008flx}}
}
\end{picture}
\ifthenelse{\boolean{captionmode}}{
\midplotoffsetcaption
\caption{The \theX seen in $\jpsi\pip\pim$ at LHCb~\cite{LHCb:2020fvo} (left), in $\Dstarz[\to\Dz\piz]\Dzb$ at BESIII~\cite{BESIII:2020nbj} (center) and in $\jpsi\g$ at BaBar~\cite{BaBar:2008flx} (right). }}{}
\midplotoffsetbottom
\end{figure}

\cardnotes{
The \theX is the first established exotic hadron and the most studied one to date.
It was discovered by Belle in 2003 in the $\jpsi\pip\pim$ system produced from \Bp decays.
The $J^{PC}$ quantum numbers of the \theX were established in analyses by CDF~\cite{CDF:2006ocq}, Belle~\cite{Belle:2011vlx} and LHCb\cite{LHCb:2013kgk,LHCb:2015jfc} to be $1^{++}$ which
makes it consistent with a $\chi_{c1}(2P)$ charmonium state. However, the observed mass disagrees with quark model calculations by around 100\mev~\cite{Barnes:2003vb} (also see Fig.~\ref{fig:specQQ}).
Moreover, it has very small width 
for a state lying above $\D\Db$ threshold.
Observation of decays to $\rho(770)J/\psi$, $\omega(782)J/\psi$ and $\pi^0 \chi_{c1}(1P)$ with comparable rates 
implies isospin violation at the level of ${\sim}0.3$, i.e. much larger than ${\sim}0.05$ seen in conventional charmonia~\cite{LHCb:2022jez}.
This is likely associated to the 
8.7\mev splitting between the $\Dz\bar{\D}^{\ast0}$ and $\Dp\D^{\ast-}$ thresholds.

Closeness to the $\Dz\bar{D}^{\ast0}$ threshold (within $-0.04\pm0.12\mev$) and prevalence of the $\Dz\bar{D}^{\ast0}$ decay mode make it a natural $\D\bar{\D}^{\ast}$ molecular candidate~\cite{Tornqvist:2004qy}. 
Prevalence of $\g\psitwos$ decay over $\g\jpsi$ observed by BaBar~\cite{BaBar:2008flx} and LHCb~\cite{LHCb:2014jvf,LHCb:2024tpv}, though disputed by Belle~\cite{Belle:2011wdj} and BESIII~\cite{BESIII:2020nbj}, 
likely indicates sizeable charmonium component in \theX~\cite{Dong:2009uf}, however, may not be in conflict with its predominantly molecular nature~\cite{Guo:2014taa}.
A pure compact tetraquark also remains viable interpretation of the \theX even though its expected charged partner is not observed.
It can be understood if production of the charged states with isospin equal to one is suppressed with respect to the isoscalar \theX~\cite{Maiani:2020zhr}.
Production properties of \theX in \pp and $\proton\bar{\proton}$ collisions are used as arguments towards presence of a compact component~\cite{Bignamini:2009sk,Esposito:2020ywk}. These arguments were, however, disputed~\cite{Artoisenet:2009wk,Braaten:2020iqw,Albaladejo:2017blx}.
}

\cardend

%% file: card_chic2p.tex
\vtop{ 

\cardstart

\cardtitle{Other $\chi_{c}$-like states below 4 GeV}{Meson-like/Hidden Charm/Isoscalar}

\begin{wrapfigure}[0]{r}{0.4\textwidth}
    \vspace{-5mm}
\ifthenelse{\boolean{captionmode}}{}{
    {\footnotesize \phantom{0} $\chi_{c0}(3915)\to\jpsi\omega$ at BaBar~\cite{BaBar:2012nxg}}
}
    \includegraphics[width=0.4\textwidth]{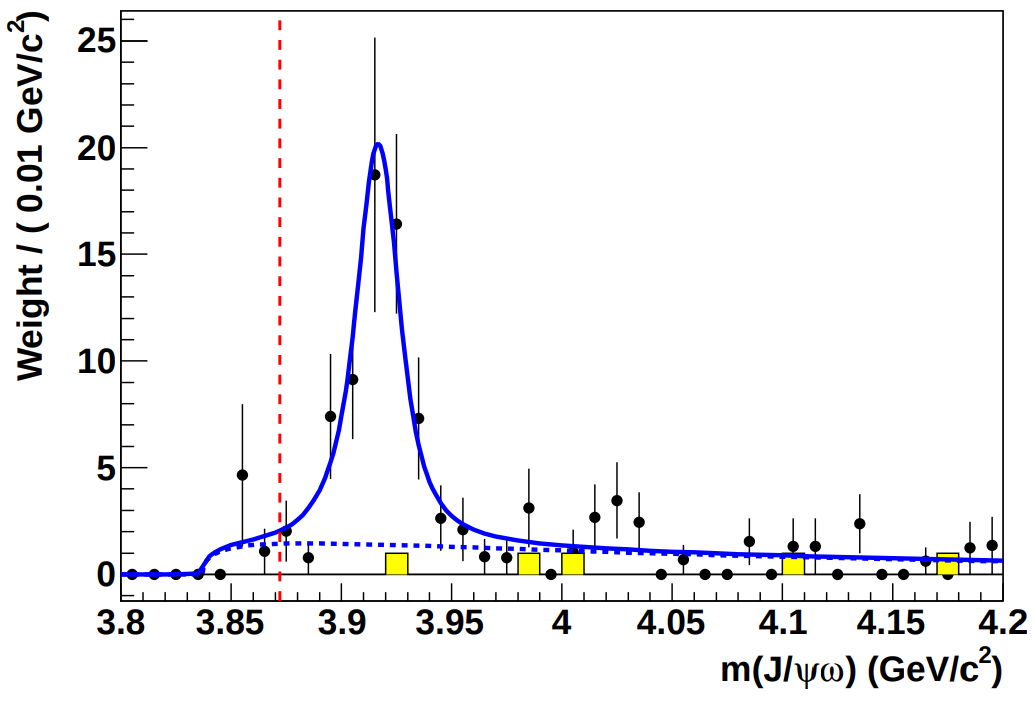}
\ifthenelse{\boolean{captionmode}}{
\wrapcaptionoffset
\caption{$\chi_{c0}(3915)\to\jpsi\omega$ signal at BaBar~\cite{BaBar:2012nxg}.}
}{}

\ifx\reallybrief 1
    \vspace{2mm}
    \includegraphics[width=0.4\textwidth]{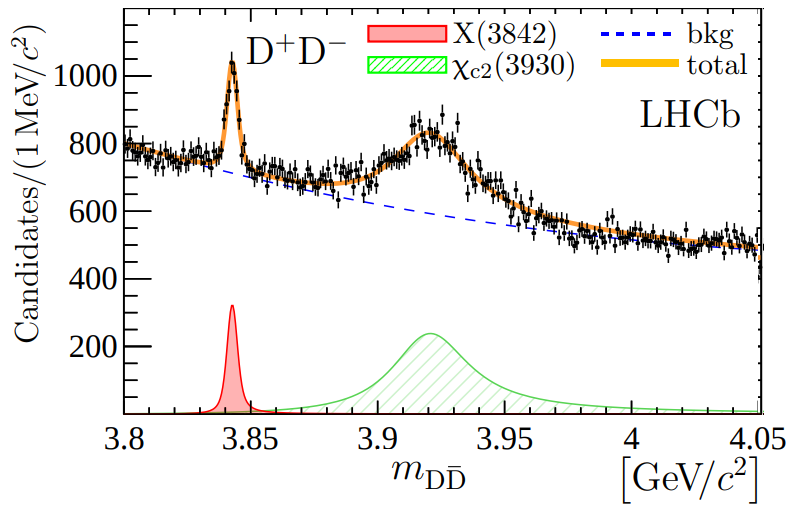}
\wrapcaptionoffset
\caption{The $\chi_{c2}(3930)\to\Dp\Dm$ at LHCb~\cite{LHCb:2019lnr}.}
\fi

\end{wrapfigure} 

\cardproperty{states}{
\cardnamelist{
 \item $I^G(J^{PC})=0^+(0^{++})$: $\chi_{c0}(3860)$, $\est{\chi_{c0}(3915)}$
 \\ \phantom{000000000000000}\cardnamealternative{\aka $X(3915)$}
 \item $I^G(J^{PC})=0^+(2^{++})$: $\est{\chi_{c2}(3930)}$
 \item $I^G(J^{PC})=?^?(?^{??})$: $X(3940)$
 }
}

\cardproperty{minimal quark content}{$[\c\cbar]$, possibly $[\c\cbar\q\qbar]$}

\cardproperty{experiments}{Belle, BaBar, BESIII, LHCb}

\cardproperty{production}{$\g\g$-collisions and 
\\ $\B$-decays ($\chi_{c0}(3915)$, $\chi_{c2}(3930)$),
\\ also $\chi_{c2}(3930)$ in \pp-collisions,
\\ $\chi_{c0}(3860)$, $X(3940)$ in $e^+e^-\to J/\psi X$,
\\ and $\chi_{c0}(3915)$ possibly in $e^+e^-\to \gamma X$
}

\cardproperty{decay modes}{$D\bar{D}$ (except $X(3940)$),
\\ $D^*\bar{D}$ ($X(3940)$),
$\omega J/\psi$ ($\chi_{c0}(3915)$)
}

\cardproperty{nearby thresholds}{$D^*\bar{D}$, $D^+_s D_s^-$}

\cardproperty{characteristic widths}{${\sim}200\mev$ ($\chi_{c0}(3860)$)
\\ and 19-37\mev (${\chi_{c0}(3915)}$, ${\chi_{c2}(3930)}$, $X(3940)$)
}

} 

\ifx\reallybrief 1
\else

\ifthenelse{\boolean{formpla}}{
\def\pictW{110}
\def\pictH{30}
}{
\def\pictW{130}
\def\pictH{38}
}

\begin{figure}[H]
\centering
\setlength{\unitlength}{1mm}
\midplotoffsettop
\begin{picture}(\pictW,\pictH)
\ifthenelse{\boolean{formpla}}{
\put(0,0){\includegraphics*[height=30mm]{"./plots/X3930_LHCb.png"}}
\put(60,0){\includegraphics*[height=30mm]{"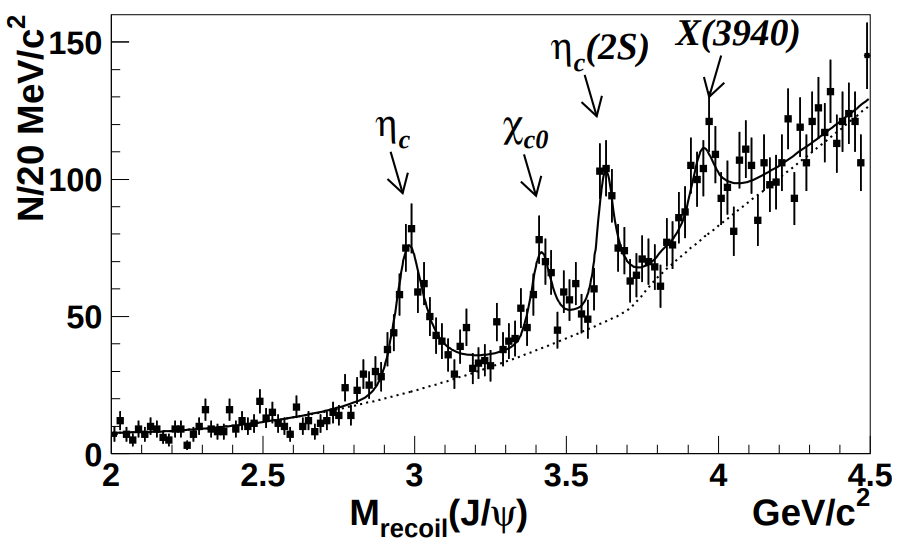"}}
}{
\put(0,0){\includegraphics*[height=38mm]{"./plots/X3930_LHCb.png"}}
\put(68,0){\includegraphics*[height=38mm]{"./plots/X3940_Belle.png"}}
}
\ifthenelse{\boolean{captionmode}}{}{
\put(5,32){\footnotesize $\chi_{c2}(3930)\to\Dp\Dm$ at LHCb~\cite{LHCb:2019lnr}}
\put(60,32){\footnotesize $X(3940)$ in $\ee\to\jpsi X$ at Belle~\cite{Belle:2005lik}}
}
\end{picture}
\ifthenelse{\boolean{captionmode}}{
\midplotoffsetcaption
\caption{The $\chi_{c2}(3930)\to\Dp\Dm$ at LHCb~\cite{LHCb:2019lnr} (left) and $X(3940)$ in $\ee\to\jpsi X$ at Belle~\cite{Belle:2005lik}.}}{}
\midplotoffsetbottom
\end{figure}
\fi

\cardnotes{The spectrum of non-vector charmonium states above the open-charm threshold is difficult to study in experiments and as such data is rather limited. Still, interesting exotic hadron candidates have emerged sharing the quantum numbers of the conventional $\chi_{cJ}$ states $J^{++}$, most prominently the $\chi_{c1}(3872)$ that we addressed in more detail in the previous card. Outside of the $\chi_{c1}(3872)$, the picture is less clear.

The Belle experiment made the first observations of a $\chi_{c0}(3915)$ and a $\chi_{c2}(3930)$ in the ${\Bp \to \omega \jpsi\Kp}$~\cite{Belle:2004lle} and $\g\g\to\D\Db$~\cite{Belle:2005rte} processes, respectively. Both observations were later confirmed by BaBar~\cite{BaBar:2007vxr,BaBar:2010jfn}.
In addition, Belle and BaBar saw a state consistent with the $\chi_{c0}(3915)$ in the two-photon production ${\g\g \to \omega \jpsi}$~\cite{Belle:2009and,BaBar:2012nxg}, with Belle also claiming evidence for a structure consistent with either the $\chi_{c0}(3915)$ or $\chi_{c2}(3930)$ in $\g\g \to \g \psi(2S)$~\cite{Belle:2021nuv}. BESIII found evidence for the $\chi_{c0}(3915)$ in ${\ee \to \g\omega \jpsi}$~\cite{BESIII:2019qvy}, although it is unclear whether a broad or a narrow structure is needed.
LHCb observed the $\chi_{c2}(3930)$ in prompt \pp production, decaying to $\D\Db$~\cite{LHCb:2019lnr}. In a recent analysis of the decay ${\Bp\to\Dp\Dm\Kp}$, they also found contributions from two states around 3.93\gev, consistent with the $\chi_{c0}(3915)$ and $\chi_{c2}(3930)$~\cite{LHCb:2020pxc}, marking the first observation of the $\chi_{c2}(3930)$ in $B$-hadron decays.
The $\chi_{c0}(3860)$ was claimed by Belle in double-charmonium production $\ee \to \jpsi \D\Db$~\cite{Belle:2017egg}. Its width is found to be much larger than the ones of the $\chi_{c0}(3915)$ and $\chi_{c2}(3930)$ (100-400\mev vs. 20-35\mev). No sign of this state was found by LHCb in ${\Bp\to\Dp\Dm\Kp}$.

The relation of these states to the radially excited charmonium triplet states $\chi_{cJ}(2P)$ is unclear (see Fig.~\ref{fig:specQQ}).
The small mass gap between the $\chi_{c0}(3915)$ and $\chi_{c2}(3930)$ as well as the strong OZI-suppressed decay of the $\chi_{c0}(3915)$ to $\omega J/\psi$ render an interpretation of those two states as the $\chi_{c0}(2P)$ and $\chi_{c2}(2P)$ somewhat unlikely. 
As such, the $\chi_{c0}(3860)$ might be a better candidate for the $\chi_{c0}(2P)$-state, but it has only ever been observed in a single channel. 
Outside of the $\chi_{c1}(3872)$ that is unlikely to be a (pure) charmonium-state, no good candidate for the $\chi_{c1}(2P)$-state exists. 
While the $X(3940)$ observed by Belle in $\ee \to \jpsi X$ with decay to $D^*\bar{D}$~\cite{Belle:2005lik} could be a candidate for the $\chi_{c1}(2P)$ state, its spin-parity has not been determined, nor has the state been confirmed by another experiment.
An assignment of $\chi_{c0}(3915)$ and $\chi_{c2}(3930)$ as regular charmonium states is further complicated by the expectation for the mass of the $\chi_{c1}(2P)$ to be in between the $\chi_{c0}(2P)$ and $\chi_{c2}(2P)$.
The situation might be clarified with observation of the $h_c(2P)$ state which will define the center-of-gravity for the spin-triplet masses. If a very recent report of a potential $h_c(2P)$ candidate with a mass around 4 GeV \cite{LHCb:2024vfz} is confirmed in the future, a conventional interpretation of the $\chi_{c0}(3915)$ and $\chi_{c2}(3930)$ would be difficult.

}

\cardend

%% file: card_psi4230.tex
\vtop{ 

\cardstart

\cardtitle{$\psi$-like states}{Meson-like/Hidden Charm/Isoscalar}

\ifthenelse{\boolean{formpla}}{ 
\def\plotwidth{0.35\textwidth}
}{ 
\def\plotwidth{0.35\textwidth}
}

\begin{wrapfigure}[0]{r}{\plotwidth}
    \vspace{-5mm}
\ifthenelse{\boolean{captionmode}}{}{
    {\footnotesize \hspace{1.5em} $\pip\pim\jpsi$ at BESIII~\cite{BESIII:2022qal}}
}
    \includegraphics[width=\plotwidth]{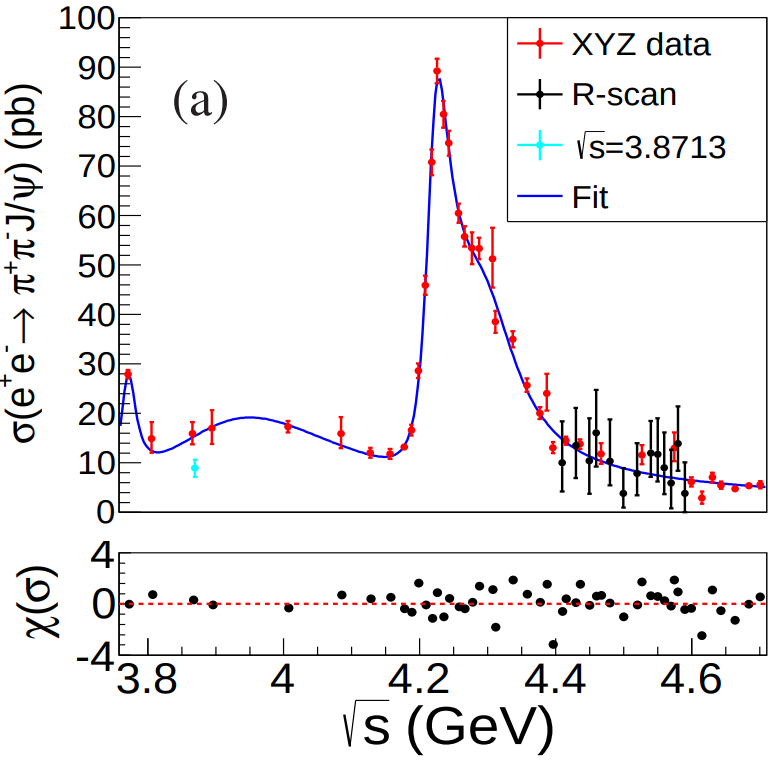}
\ifthenelse{\boolean{captionmode}}{
\wrapcaptionoffset
\caption{$\ee\to\pip\pim\jpsi$ cross-section at BESIII~\cite{BESIII:2022qal}.}
}{}
\end{wrapfigure} 

\cardproperty{states}{
$\est{\psi(4230)}$, $\est{\psi(4360)}$, $\psi(4660)$

 \cardnamealternative{\aka $Y(4230)$, $\psi(4260)$, $Y(4360)$, ...
}
}

\cardproperty{quantum numbers}{$I^G(J^{PC}) = 0^-(1^{--})$}

\cardproperty{minimal quark content}{$[\c\cbar]$, \\ possibly $[\c\cbar\q\qbar]$ or $[\c\cbar g]$}

\cardproperty{experiments}{{BaBar}, CLEO, Belle, BESIII, \\ possibly D0}

\cardproperty{production}{$e^+e^-$ annihilation,\\ possibly $\b$-decays ($\psi(4230)$)}

\cardproperty{decay modes}{$\pi\pi J/\psi$, $\pi\pi \psi(2S)$, $\pi\pi h_c$ \\ (possibly via $\pi T_{c\bar{c}}$) and $\eta^{(\prime)}\jpsi$ for $\psi(4230|4360)$,
\\ also $K\bar{K} J/\psi$, $3\pi\eta_c$, $\omega\chi_{c0}$, $\gamma \chi_{c1}(3872)$, $\mu^+\mu^-$,  
\\ $D^\ast\bar{D}\pi$, potentially $D\bar{D}$ for $\psi(4230)$,
\\ $\pi\pi\psi_2(3823)$, $D^+D^-\pi^+\pi^-$, 
\\ possibly $\pi\pi\psi(3770)$ and $\D_{1}(2420)\bar{\D}$ for $\psi(4360)$, 
\\ $\pi\pi \psi(2S)$, possibly $\Lambda_c \bar{\Lambda}_c$ for $\psi(4660)$
}

\cardproperty{nearby thresholds}{$D_1\bar{D}$, $D_s^{*+}D_s^{*-}$}

\cardproperty{characteristic widths}{48-118\mev}

} 

\ifx\reallybrief 1
\else
\ifthenelse{\boolean{formpla}}{
\def\pictW{120}
\def\pictH{30}
}{
\def\pictW{150}
\def\pictH{38}
}

\begin{figure}[H]
\centering
\setlength{\unitlength}{1mm}
\midplotoffsettop
\begin{picture}(\pictW,\pictH)

\ifthenelse{\boolean{formpla}}{
\put(0,0){\includegraphics*[height=30mm]{"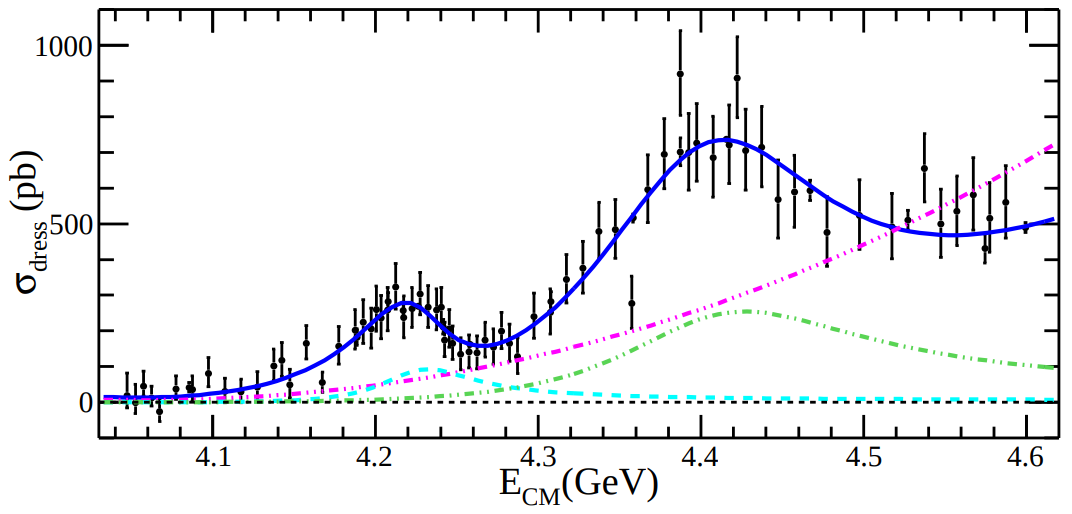"}}
\put(70,0){\includegraphics*[height=30mm]{"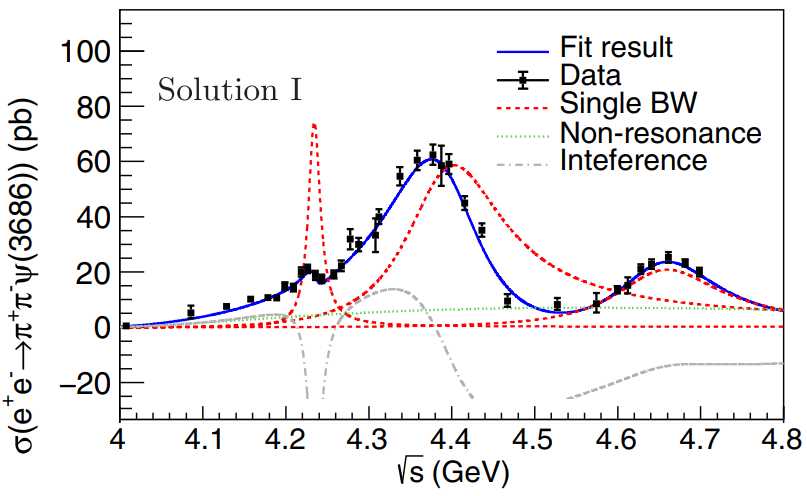"}}
}{
\put(0,0){\includegraphics*[height=40mm]{"./plots/DDstpi_psi-like_BESIII.png"}}
\put(85,0){\includegraphics*[height=40mm]{"./plots/pipipsi2S_BESIII_cropped.png"}}
}

\ifthenelse{\boolean{captionmode}}{}{
\put(10,32){\footnotesize $\pip\Dz\D^{\ast-}$ at BESIII~\cite{BESIII:2018iea}}
\put(75,32){\footnotesize $\pip\pim\psitwos$ at BESIII~\cite{BESIII:2021njb}}
}
\end{picture}
\ifthenelse{\boolean{captionmode}}{
\midplotoffsetcaption
\caption{Exclusive $\ee\to\pip\Dz\D^{\ast-}$ (left)
and $\ee\to\pip\pim\psitwos$ (right) cross-sections measured by BESIII~\cite{BESIII:2018iea,BESIII:2021njb}.
}}{}
\midplotoffsetbottom
\end{figure}
\fi

\cardnotes{ 

The $Y(4260)$, first observed as a broad peak in the process $e^+e^-\to \gamma_{\textrm{ISR}}\pi\pi J/\psi$ by BaBar~\cite{BaBar:2005hhc}, was later resolved into two distinct structures by BESIII~\cite{BESIII:2016bnd}, now called $\psi(4230)$ and $\psi(4360)$. 
In fact, the latter was already observed in the process $e^+e^-\to \gamma_{\textrm{ISR}}\pi\pi\psi(2S)$ by BaBar~\cite{BaBar:2006ait} and Belle~\cite{Belle:2007umv}.
The spin-parity is unambiguously fixed through the production process. Today, many processes of the type $e^+e^-\to (c\bar{c}) (q\bar{q})$ are observed to have large structures in the region between 4.2 and 4.4\gev, that are commonly attributed to these two states. In case of the processes $e^+e^-\to J/\psi \pi\pi$, $h_c\pi\pi$ and $\psi(2S)\pi\pi$, intermediate $T_{\c\cbar}$-states appear -- often for a specific range of center-of-mass energy, suggesting decays of the type $\psi(4230)$ or $\psi(4360)\to T_{\c\cbar}\pi$.
However, resonance parameters extracted in individual final states are often inconsistent, leaving some ambiguity on the number of states observed in this mass region.
The only hint for the $\psi(4230)$ outside of $e^+e^-$ annihilation stems from D0, who claims a $T_{\c\cbar1}(3900)$ signal in inclusive $b$-decays to $J/\psi \pi^+\pi^-$, but only if the $J/\psi \pi^+\pi^-$ system has invariant mass in the range between $4.2$ and $4.3 \gev$~\cite{D0:2018wyb, D0:2019zpb}. This claim has yet to be confirmed by LHCb or Belle(II).

If the $\psi(4040)$, $\psi(4160)$ and $\psi(4415)$ are identified with the $\psi(3S)$, $\psi(2D)$ and $\psi(4S)$ states, respectively, the $\psi(4230)$ and $\psi(4360)$ can hardly be explained as conventional charmonia~\cite{Hanhart:2019isz}.
This is further supported by the fact that $\psi(4230)$ and $\psi(4360)$ are seen most prominently in OZI-suppressed decays to hidden-charm channels, although it should be noted that interpretation of open-charm cross sections is more difficult and decays to these OZI-favored channels are not ruled out.
Given the surprisingly few observations of the $\psi(4040)$, $\psi(4160)$ and $\psi(4415)$ in exclusive processes, a better understanding of all vector-states above the open-charm threshold is required to claim the existence of exotic hadrons with $J^{PC}=1^{--}$. Coupled channel analyses \cite{Eichten:1978tg,Eichten:1979ms,Cao:2014qna,Du:2016qcr,Uglov:2016orr,Nakamura:2023obk,Husken:2024hmi} might help improve this situation.
The $\psi(4230)$ might be a key state to understand the bigger picture of exotic charmonium-like hadrons, given that its decays to the $\pi T_{c\bar{c}1}(3900)$ and $\g\chi_{c1}(3872)$ clearly suggest a common explanation to all three exotic hadron candidates. One such explanation that nicely ties these decays together is the $\psi(4230)$ as a $D_1(2420)\bar{D}$ molecule~\cite{Wang:2013cya,Guo:2013zbw}. 

Structures that appear in \ee annihilation at higher energies are usually attributed to one state -- $\psi(4660)$. 
It is, for example, seen in $\pip\pim\psi(2S)$ by Belle~\cite{Belle:2007umv}, BaBar~\cite{BaBar:2012hpr} and BESIII~\cite{BESIII:2021njb}. A near-threshold peak in $\Lambda_c^+ \Lambda_c^-$ seen by Belle~\cite{Belle:2008xmh} is, however, not confirmed by BESIII~\cite{BESIII:2017kqg}. Recent BESIII data on $K\bar{K}\jpsi$~\cite{BESIII:2023wqy} and $\D_s^{\ast+}\D_s^{\ast-}$~\cite{BESIII:2023wsc} in addition indicate structures around 4.5 and 4.7\gev.}

\cardend

%% file: card_JpsiPhi.tex
\vtop{ 

\cardstart

\cardtitle{ 
States seen in $\jpsi\phi$}{Meson-like/Hidden Charm/Isoscalar}

\begin{wrapfigure}[0]{r}{0.4\textwidth}
    \vspace{-7mm}
\ifthenelse{\boolean{captionmode}}{}{
    {\color{black} \footnotesize \phantom{000} Resonances in $\jpsi\phi$ at LHCb~\cite{PhysRevLett.118.022003}} \\
}
    \includegraphics[width=0.43\textwidth]{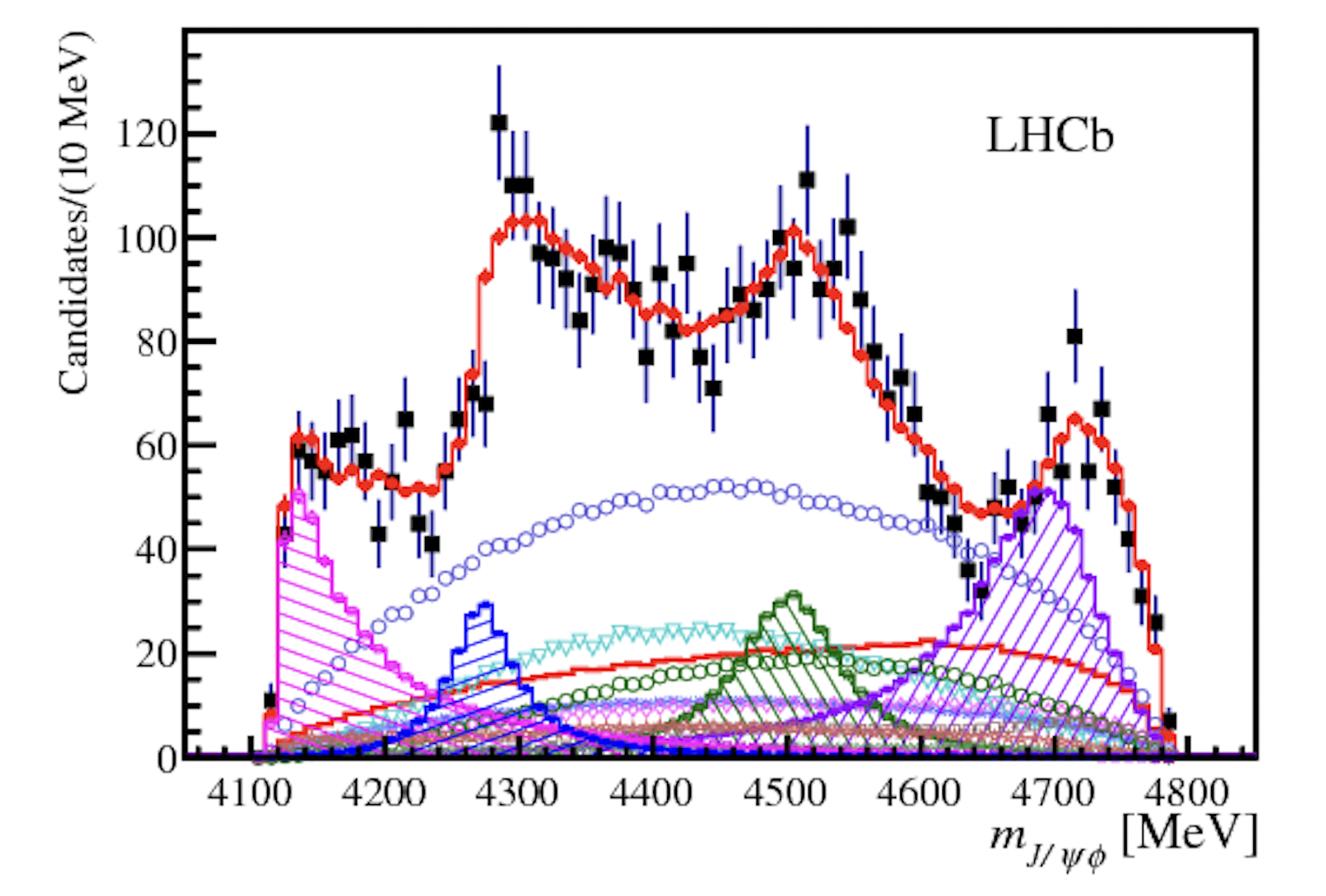}
\ifthenelse{\boolean{captionmode}}{
\wrapcaptionoffset
\caption{Resonances in $\jpsi\phi$ at LHCb~\cite{PhysRevLett.118.022003}.}
}{}
\end{wrapfigure} 

\cardproperty{states}{
    \cardnamelist{
    \item $I(J^{PC})=0(0^{++})$:  $\est{\chi_{c0}(4500)}$, $\chi_{c0}(4700)$
    
    \item $I(J^{PC})=0(1^{++})$: $\est{\chi_{c1}(4140)}$, $\est{\chi_{c1}(4274)}$, \\ $\chi_{c1}(4685)$ \\
    \cardnamealternative{\aka $X(4140)$, $Y(4140)$, ...}
    \item $I(J^{PC})=0(?^{?+})$: $X(4150)$, $X(4630)$, $X(4740)$ \\
    \cardnamealternative{\aka $X(4160)$}
    }
}

\cardproperty{minimal quark content}{$[\c\cbar]$, 
\\ more likely $[\c\cbar\q\qbar]$ or $[\c\cbar\s\sbar]$
}

\cardproperty{experiments}{CDF, CMS, D0, BaBar,\\ LHCb, Belle
}

\cardproperty{production}{$\Bp\to\jpsi\phi \Kp$, 
\\ $\Bs\to\jpsi\phi\pip\pim$ ($X(4740)$), 
\\ exclusive $\proton\proton\to\proton\proton+\jpsi\phi$ ($\chi_{c1}(4274)$, $\chi_{c0}(4500)$),
\\ possibly $p\bar{p}\to\jpsi\phi+X$ ($\chi_{c1}(4140)$), 
\\ $e^+ e^- \to \jpsi \D^{\ast}\bar{\D}^{\ast}$ ($X(4150)$) }

\cardproperty{decay modes}{$\jpsi \phi$, $\D^{\ast}\bar{\D}^{\ast}$ ($X(4150)$)}

\cardproperty{nearby threshold}{$\D^{}_s\bar{\D}_s^{\ast}$, $\D_s^{\ast}\bar{\D}_s^{\ast}$ 
}

\cardproperty{characteristic widths}{51-174 \mev \footnotemark
}

} 

\footnotetext{for the width of the $\chi_{c1}(4140)$ state we consider only LHCb measurements as all the others are one-dimensional and hence are likely biased, ignoring interference with other decay chains
}

\cardnotes{
In 2009, the CDF experiment found a near-threshold structure in the $\jpsi\phi$ system with a mass around $4140\mev$ in the decay $\Bp\to\jpsi\phi \Kp$~\cite{PhysRevLett.102.242002}, the $\chi_{c1}(4140)$.
Similar enhancements were later confirmed by CDF~\cite{CDF:2011pep}, CMS~\cite{PhysLetB.734.261-281}, and D0~\cite{D0:2013jvp}, alongside a potential second state around $4270\mev$. D0 later also found a similar structure in prompt production in $p\bar p$ collisons~\cite{D0:2015nxw}.
Hints for both states were also seen by~BaBar~\cite{BaBar:2014wwp}.
However, masses and widths of the two structures identified in these simple analyses of one-dimensional mass 
distributions were found to disagree with each other.
In turn, the Belle~\cite{Cheng-Ping:2009sgk,Belle:2009rkh}, BESIII~\cite{BESIII:2014fob} and LHCb~\cite{LHCb:2012wyi} experiments did not find the $\chi_{c1}(4140)$ at that time.
Only later, the amplitude analysis of  high-statistics $\Bp\to\jpsi\phi \Kp$ decays by LHCb~\cite{PhysRevLett.118.022003, PhysRevLett.127.082001} recovered both the $\chi_{c1}(4140)$ and the $\chi_{c1}(4274)$ state, however, with substantially different masses and widths. The quantum numbers of both states were determined.
In the same analysis LHCb finds four additional states as well as evidence for a fifth one: 
the $\chi_{c0}(4500)$ and $\chi_{c0}(4700)$ with $J^{PC}=0^{++}$, and the $X(4630)$ ($J^{PC} = 1^{-+}$ or $2^{-+}$) and $\chi_{c1}(4685)$ ($J^{PC} = 1^{++}$), with evidence for the
$X(4150)$, previously observed by Belle in $e^+e^- \to \jpsi \Dstar \bar{\D^*}$~\cite{Belle:2007woe} at the 4.8$\sigma$ level, with a preferred spin-parity of $J^P = 2^-$. 
In addition, a potential new structure labelled as $X(4740)$ is observed  by LHCb in the decay $\Bs\to\jpsi\phi\pip\pim$~\cite{LHCb:2020coc} from the $\jpsi\phi$ invariant mass distribution. An amplitude analysis with larger statistics is needed to fully establish whether the enhancement can be identified with the $\chi_{c0}(4700)$ from $\Bp\to\jpsi\phi \Kp$. 
Recently, LHCb observed both $\chi_{c1}(4274)$ and $\chi_{c0}(4500)$ states (together with hints for more) in $\proton\proton\to\proton\proton+\jpsi\phi$ collisions with no additional activity~\cite{LHCb:2024smc}.

Some of these states may be considered as candidates for conventional charmonia: the $\chi_{c1}(4140)$
or the $\chi_{c1}(4274)$ are candidates for the $\chi_{c1}(3P)$ state; the $\chi_{c0}(4500)$ and the $\chi_{c0}(4700)$ could correspond to the $\chi_{c0}(4P)$ and $\chi_{c0}(5P)$. 
In that case they should decay to $\jpsi\omega$ with comparable rates to the $\jpsi\phi$. This is yet to be explored.
While $\D_s^{*+}\D_s^{*-}$ molecular states with $J^{PC}=0^{++}$ or $2^{++}$ have been suggested~\cite{PhysRevD.80.017502} as an interpretation of the $\chi_{c1}(4140)$, the measured spin-parity $J^P=1^+$ rules that out.
Alternatively, it could be a $\D_s^{+}\D_s^{\ast-}$ molecule bound by an $\eta$ meson exchange~\cite{Karliner_2016}, or a corresponding cusp~\cite{PhysRevD.80.114013,LHCb:2016nsl}. To clarify this question, exploration of the $\D_{\s}^{+}\D_{\s}^{(\ast)-}$ decay mode is needed.

With the coming energy-upgrade to the BEPCII accelerator, the high-mass structures in $J/\psi \phi$ could become accessible to BESIII, allowing for an independent confirmation if production in e.g. $e^+e^-\to \gamma J/\psi \phi$ is strong enough.
} 

\cardend

%% file: card_Zc3900.tex
\vtop{ 

\cardstart

\cardtitle{$T_{\c\cbar}$ states seen in $e^+e^-$ annihilation
\cardnamealternative{\hspace{-1.5em}(\aka $Z_{\c}$, $X$, $T_{\psi}$)}
}{Meson-like/Hidden Charm/Isovector}

\begin{wrapfigure}[0]{r}{0.4\textwidth}
\ifthenelse{\boolean{captionmode}}{}{
    {\color{black} \footnotesize \phantom{000} ${T_{\c\cbar1}(3900)^{+/0}}$ at BESIII ??? \todo{Reference}{} } \\
}
    \includegraphics[width=0.4\textwidth]{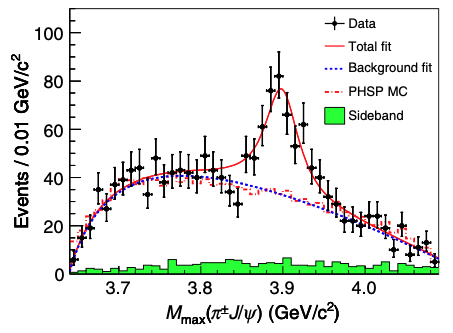}
\ifthenelse{\boolean{captionmode}}{
\wrapcaptionoffset
\caption{${T_{\c\cbar1}(3900)}$ signal at \mbox{BESIII}~\cite{BESIII:2013ris}.}
}{}
\end{wrapfigure}

\cardproperty{states}{
    \cardnamelist{
    \item $I^G(J^{PC})=1^+(1^{+-})$:  $\est{T_{\c\cbar1}(3900)^{+/0}}$
    \item $I^G(J^{PC})=1^+(?^{?-})$: $\est{T_{\c\cbar}(4020)^{+/0}}$, $T_{\c\cbar}(4055)^+$
    }
}

\cardproperty{minimal quark content}{$[c\bar{c}\q\qbar']$}

\cardproperty{experiments}{Belle, BESIII, CLEO-c, \\ possibly D0}

\cardproperty{production}{$e^+e^-\to T_{c\bar c}\pi$ \\ possibly through $\psi(4230)/\psi(4360)$, \\
$T_{\c\cbar1}(3900)$ possibly in $\b\to\jpsi\pip\pim{+X}$
}

\cardproperty{decay modes}{$\pi J/\psi$, $D^\ast \bar{D}$ and \\ possibly $\rho \eta_c$ ($T_{c\cbar1}(3900)$),
\\ $\pion h_c(1P)$, $D^\ast \bar{D}^\ast$ ($T_{c\cbar}(4020)$),
\\ $\pip\psitwos$ ($T_{\c\cbar}(4055)$)
}

\cardproperty{nearby thresholds}{$D^{\ast}\bar{D}$, $D^{\ast}\bar{D}^\ast$}

\cardproperty{characteristic widths}{13-45 MeV}

} 

\cardnotes{

The $T_{c\bar{c}1}(3900)^\pm$ was first observed simultaneously by Belle and BESIII in $e^+e^-\to T_{c\bar{c}1}(3900)^\pm \pi^\mp$~\cite{Belle:2013yex,BESIII:2013ris}, with Belle using the initial-state-radiation technique. It was confirmed by CLEO-c soon after, with CLEO-c also claiming first evidence for the neutral partner state~\cite{Xiao:2013iha}, now confirmed by BESIII with much larger statistics~\cite{BESIII:2015cld,BESIII:2020oph}. A spin-parity of $J^{P(C)}=1^{+(-)}$ was established by BESIII in an amplitude analysis of the process $e^+e^- \to J/\psi \pi^+\pi^-$~\cite{BESIII:2017bua} (where only the neutral state is an eigenstate to $C$). Its mass is close to the $D^\ast \bar{D}$ threshold, which is also its dominant decay mode~\cite{BESIII:2013qmu,BESIII:2015ntl}.
As an isospin-triplet, the $T_{c\bar{c}1}(3900)$ is openly exotic, clearly requiring at least a $c\bar{c}$- and a light-quark pair. 
A similar state $T_{c\bar{c}}(4020)$ is found near the $D^\ast \bar{D}^\ast$ threshold, decaying to $h_c\pi$~\cite{BESIII:2013ouc,BESIII:2014gnk} and $D^\ast \bar{D}^\ast$~\cite{BESIII:2013mhi,BESIII:2015tix}, that is supposedly related to the $T_{c\bar{c}1}(3900)$.

These $T_{c\bar c}$ states are largely interpreted as molecular candidates, although interpretation as compact diquark anti-diquark states have not been fully ruled out, and discussions about triangle singularities persist - see Refs.~\cite{Maiani:2013nmn,Wang:2013cya,Liu:2013vfa,Szczepaniak:2015eza,Pilloni:2016obd,Albaladejo:2015lob,Gong:2016jzb,Guo:2019twa,Ding:2020dio}. So far, all observations of the $T_{c\bar{c}1}(3900)$ are related to signals around 4.2 to 4.3 GeV in $T_{c\bar{c}1}(3900) \pi$, which can be explained in the molecular picture through de-excitation of a $D_1 \bar{D}$ molecule to a $D^\ast \bar{D}$ with pion emission.  Noticeably, $T_{c\bar{c}}$-like states observed in three-body final states in $e^+e^-$ annihilation are inconsistent with those seen in three-body $B$-decays. Alternative explanations related to triangle singularities can in the future be investigated in an independent third production process, such as photo-production experiments at JLab and the EIC using $\gamma p \to T_{c\bar c}^+ n$ or $\gamma p \to T_{c\bar c}^- \Delta^{++} $~\cite{AbdulKhalek:2021gbh,Winney:2022tky,Accardi:2023chb}. }

\cardend

%% file: card_Zc_fromB.tex
\vtop{ 

\cardstart

\cardtitle{$T_{\c\cbar}$ states seen in \b-hadron decays
\cardnamealternative{\hspace{-1.5em}(\aka $Z_{\c}$, $X$, $T_{\psi}$)}
}{Meson-like/Hidden Charm/Isovector}

\begin{wrapfigure}[0]{r}{0.37\textwidth}
\ifthenelse{\boolean{captionmode}}{}{
    {\color{black} \footnotesize \hspace{4em} $T_{\c\cbar1}(4430)^+$ at LHCb~\cite{LHCb:PhysRevLett.112.222002}} \\
}
    \includegraphics[width=0.37\textwidth]{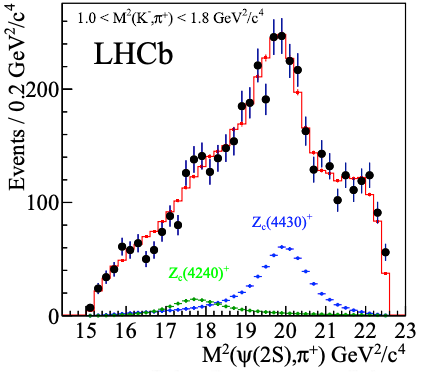}
\ifthenelse{\boolean{captionmode}}{
\wrapcaptionoffset
\caption{$T_{\c\cbar1}(4430)^+$ signal at LHCb~\cite{LHCb:PhysRevLett.112.222002}.}
}{}
\end{wrapfigure}

\cardproperty{states}{
    \cardnamelist{
    \item $I^G(J^{PC}) = 1^+(1^{+-})$: $T_{c\bar{c}1}(4200)^+$, $\est{T_{\c\cbar1}(4430)^+}$
    \item $I^G(J^{PC}) = 1^+(?^{?-})$: $T_{\c\cbar}(4240)^+$ 
    \\ \cardnamealternative{\aka $R_{c0}(4240)$, $Z_{c}(4240)$}
    \item $I^G(J^{PC}) = 1^-(?^{?+})$: 
    $T_{c\bar{c}}(4050)^+$, $T_{\c\cbar}(4100)^+$, \\ $T_{c\bar{c}}(4250)^+$
    }
}

\cardproperty{minimal quark content}{$[c\bar{c}\q\qbar']$}

\cardproperty{experiments}{Belle, LHCb}

\cardproperty{production}{$\bar{\B}^0\to (c\bar c)\pip \Km$, where \\ $(c\bar c) = \jpsi$, \psitwos, $\eta_c$, $\chi_{c1}$ \\
$T_{\c\cbar}(4200)$ also potentially in $\Lambda_b\to \jpsi\pi^- p$ 
}

\cardproperty{decay modes}{$\jpsi\pip$, $\psitwos\pip$, $\eta_{c}\pip$, $\chi_{c1}\pip$
}

\cardproperty{nearby threshold}{$D^{\ast}\bar{D}^\ast$}

\cardproperty{characteristic widths}{82-370 MeV}

} 

\cardnotes{

As the first charged charmonium-like state with manifestly exotic nature, the $T_{\c\cbar1}(4430)^+$, was observed by Belle in 2007 in the decays $\bar{\B}\to\psitwos\pip \bar{K}$ (with $\bar{K} = K_s^0$ or $\Km$) as a peaking structure in the $\psitwos\pip$ invariant mass~\cite{PhysRevLett.100.142001}. Results were later updated with a full amplitude analysis which favored a spin-parity of $1^+$ at 3.4$\sigma$~\cite{Belle:PhysRevD.88.074026}.  
Independent confirmation was made by LHCb~\cite{LHCb:PhysRevLett.112.222002} with a four-dimensional analysis of the same decays but with an order of magnitude larger statistics than Belle. LHCb confirmed the Belle results for the mass, a width of about 180 MeV, and a spin-parity of $1^+$. 
A resonant interpretation is supported by clear phase-motion in the Argand diagram. 
In the same analysis, a second resonance was observed with a significance of 6$\sigma$ at a mass of $4240$ MeV, and an even larger width of $220$ MeV, called the $T_{\c\cbar}(4240)^+$. Its preferred quantum numbers are $J^P=0^-$. 
Though preference over $1^+$ is only on the level of $1\sigma$, a $1^+$ assignment requires a width of $660\pm150\mev$ which we consider less plausible.
In an amplitude analysis of $\bar{\B}^0\to\jpsi\pip\Km$ decays Belle observed another structure
at a mass of 4200\mev, called the $T_{\c\cbar1}(4200)^+$~\cite{Belle:PhysRevD.90.112009}. 
This structure, with a very broad width of about 370 MeV, interferes with the $T_{\c\cbar1}(4430)^+$ producing a dip in the $\jpsi\pip$ invariant mass at the $T_{\c\cbar1}(4430)^+$ position. 

Several additional claims have been also reported needing independent confirmation.
Two structures were observed in $\bar{\B}^0\to\chi_{c1}\pip \Km$ in the $\chi_{c1}\pip$ system by Belle, called the $T_{\c\cbar}(4050)^+$ and
$T_{\c\cbar}(4250)^+$ using a two-dimensional Dalitz plot analysis~\cite{Belle:PhysRevD.78.072004}. 
Another structure is found in $\eta_c \pi$ at around $4100$ MeV by LHCb in the decay $\B^0\to\eta_c\pim \Kp$~\cite{LHCb:2018oeg}
with a significance just above 3$\sigma$.
If confirmed, the $T_{\c\cbar}(4100)$ would be the first heavy-quark exotic hadron found decaying to two pseudoscalars. 
The favoured quantum number assignments are $J^P=0^+$ or $J^P=1^-$ and currently can not be discriminated.

Most of the structures observed in these three-body $B$-decays have been linked to triangle singularities in the past (see Refs.~\cite{Nakamura:2019btl,Nakamura:2019emd}). Even though the typical phase-motion of a resonance has been established for the $T_{\c\cbar1}(4430)^+$~\cite{LHCb:PhysRevLett.112.222002}, it is argued that triangle singularities are also able to produce that signature. At present, drawing firm conclusions on the resonant nature of these states thus appears difficult, with clarification hopefully provided by future measurements in photo-production.
}

\cardend

%% file: card_Zcs.tex
\vtop{ 

\cardstart

\cardtitle{$T_{\c\cbar\s}$ states
\cardnamealternative{(\aka $T_{\psi\s}$ or $Z_{\c\s}$)}
}{Meson-like/Hidden Charm/Isospin=\nicefrac{1}{2}}

\begin{wrapfigure}[0]{r}{0.4\textwidth}
\vspace{-10mm}
\ifthenelse{\boolean{captionmode}}{}{
    {\color{black} \footnotesize \hspace{4em} $T_{\c\cbar\s}(4000)$ at LHCb~\todo{Reference}{}} \\
}
    \includegraphics[width=0.4\textwidth]{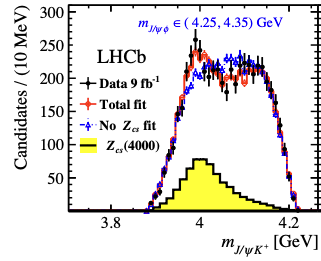}
\ifthenelse{\boolean{captionmode}}{
\wrapcaptionoffset
\caption{$T_{\c\cbar\s}(4000)$ signal at LHCb~\cite{PhysRevLett.127.082001}.}
}{}
\end{wrapfigure}  

\cardproperty{states}{ 
\cardnamelist{
\item $I(J^{P}) = \nicefrac{1}{2}(?^?)$: $T_{\c\cbar\s}(3985)^{-/0}$  
\item $I(J^{P}) = \nicefrac{1}{2}(1^{+})$: $T_{\c\cbar\s1}(4000)^{-/0}$
\item $I(J^{P}) = \nicefrac{1}{2}(1^{?})$:  $T_{\c\cbar\s1}(4220)^-$
}
}
\cardproperty{minimal quark content}{$[c\bar{c}\s\qbar]$}

\cardproperty{experiments}{{BESIII}, LHCb}

\cardproperty{production}{$e^+e^- \to T_{\c\cbar\s}(3985) \bar K$,  
\\ $\Bm\to\jpsi\phi \Km$ ($T_{\c\cbar\s1}(4000)$, $T_{\c\cbar\s1}(4220)$), \\ $\B^0\to\jpsi \phi K_S^0$ ($T_{\c\cbar\s1}(4000)$)
}

\cardproperty{decay modes}{$\D_s^-\D^{\ast\pluszero}/\D_s^{\ast -}\D^{\pluszero}$, \\
$\jpsi\Km$, $\jpsi K_S^0$ }

\cardproperty{nearby thresholds}{$\D_s^{-}\D^{\ast\pluszero}$, $\D_s^{\ast -}\D^{(\ast)\pluszero}$
}

\cardproperty{characteristic widths}{8-13\mev ($T_{\c\cbar\s}(3985)$), \\ 130-233\mev ($T_{\c\cbar\s1}(4000)$, $T_{\c\cbar\s1}(4220)$)
}

} 

\cardnotes{
First candidates for open-strange hidden-charm states are observed in the reactions $e^+e^- \to D^\ast \bar{D}_s K$ and $D \bar{D}^\ast_s K$ by the BESIII~\cite{BESIII:2020qkh} experiment and in $\B\to J/\psi \phi K$ decays by LHCb~\cite{PhysRevLett.127.082001}.  BESIII observed an enhancement in $e^+e^-\to (\D_s^-\D^{\ast 0} \Kp + \D_s^{\ast-}\D^{0} \Kp)$ near the $\D_s^-\D^{\ast 0}$ and $\D_s^{\ast -}\D^{0}$ thresholds in the $\Kp$ recoil-mass spectrum~\cite{BESIII:2020qkh}. The enhancement, called $T_{\c\cbar\s}(3985)^-$, is compatible with a $\D_s^-\D^{\ast 0}$ and $\D_s^{\ast -}\D^{0}$ resonant structure with a width of $13 \mev$, and is fitted with a mass-dependent Breit-Wigner line shape. In the one-dimensional analysis, interference effects are neglected and quantum numbers can not be established. 
Spin-parity is assumed to be $1^+$, considering that both production and decay processes occur in the favoured $S$-wave. However, other possible quantum numbers are allowed. 
Evidence for a similar (now neutral) near-threshold structure is seen in $e^+e^-\to (\D_s^+\D^{\ast -} K_S^0 + \D_s^{\ast+}\D^{-} K_S^0)$~\cite{BESIII:2022qzr}, solidifying the case for the 
enhancement attributed to the $T_{\c\cbar\s}(3985)^-$, albeit without necessarily requiring a resonant interpretation.

Signals of tetraquark candidates with strangeness are observed in $\Bm\to\jpsi\phi\Km$ decays by LHCb, where two states, $T_{\c\cbar\s1}(4000)^{-}$ and $T_{\c\cbar\s1}(4220)^-$, are found with a significance larger than 5$\sigma$. Their widths are found to be fairly broad.
Thanks to the amplitude analysis, quantum numbers are determined to be $J^P=1^+$ for the $T_{\c\cbar\s1}(4000)^{-}$ state, while $1^+$ is favored over $1^-$ for the $T_{\c\cbar\s1}(4220)^{-}$. 
An isospin partner of the $T_{\c\cbar\s1}(4000)^{-}$ was also seen in $\B^0\to\jpsi\phi K_S^0$ decays by LHCb with 4$\sigma$
significance~\cite{LHCb:2023hxg}.
Values of mass and width are consistent with the charged partner.

The two states, $T_{\c\cbar\s}(3985)$ and $T_{\c\cbar\s1}(4000)$, have compatible masses but incompatible widths, being around 10\mev and 130\mev in the two cases, which could either indicate the existence of two separate states or possibly be explained in a coupled-channel model~\cite{Ortega:2021enc}. According to the diquark model~\cite{maiani2021new}, 
they should be different but mass-degenerate states with $J^{PC}=1^{++}$ and $1^{+-}$, respectively.
} 

\cardend

%% file: card_upsilon_3states.tex
\vtop{ 

\cardstart

\cardtitle{$\Upsilon$-like states}{Meson-like/Hidden Bottom/Isoscalar}

\ifx\reallybrief 1
\begin{wrapfigure}[0]{r}{0.4\textwidth}
    \vspace{-14mm}
    \includegraphics[width=0.4\textwidth]{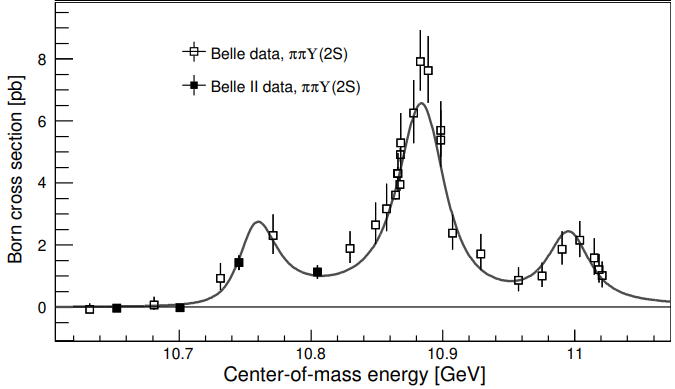}
\wrapcaptionoffset
\caption{$\ee\to\Upsilon(2S)\pip\pim$ cross-section measured by Belle and \BelleTwo~\cite{Belle-II:2024mjm}.}
\end{wrapfigure}
\fi

\cardproperty{states}{$\est{\Upsilon(10753)}$, $\est{\Upsilon(10860)}$, $\est{\Upsilon(11020)}$
\\ \cardnamealternative{\aka $\Upsilon(5S)$, $\Upsilon(6S)$}
}

\cardproperty{quantum numbers}{$I^G(J^{PC}) = 0^-(1^{--})$}

\cardproperty{minimal quark content}{$[\b\bbar]$, possibly $[\b\bbar\q\qbar]$ or $[\b\bbar g]$}

\cardproperty{experiments}{CUSB, CLEO, BaBar, {Belle}, \BelleTwo}

\cardproperty{production}{\ee annihilation}

\cardproperty{decay modes}{all in $\pip\pim \Upsilon(nS)$ ($n=1,2,3$), \\
also $\omega\chi_{b1,2}(1P)$ for $\Upsilon(10753)$ (and possibly $\Upsilon(10860)$), \\
$\pip\pim h_{b}(nP)$ ($n=1,2$), and possibly $\pip\pim\piz\chi_{\b1,2}(1P)$
}

\cardproperty{nearby thresholds}{$\B^{}_{\s}\bar{B}_{\s}$, $\B^{}_{\s}\bar{B}_{\s}^{\ast}$, $\B_{\s}^{\ast}\bar{B}_{\s}^{\ast}$}

\cardproperty{characteristic widths}{24-37\mev}

\ifx\reallybrief 1
\else
\ifthenelse{\boolean{formpla}}{
\def\pictW{125}
\def\pictH{37}
}{
\def\pictW{150}
\def\pictH{45}
}

\begin{figure}[H]
\centering
\setlength{\unitlength}{1mm}
\midplotoffsettop
\begin{picture}(\pictW,\pictH)
\ifthenelse{\boolean{formpla}}{
\put(0,0){\includegraphics*[height=37mm]{"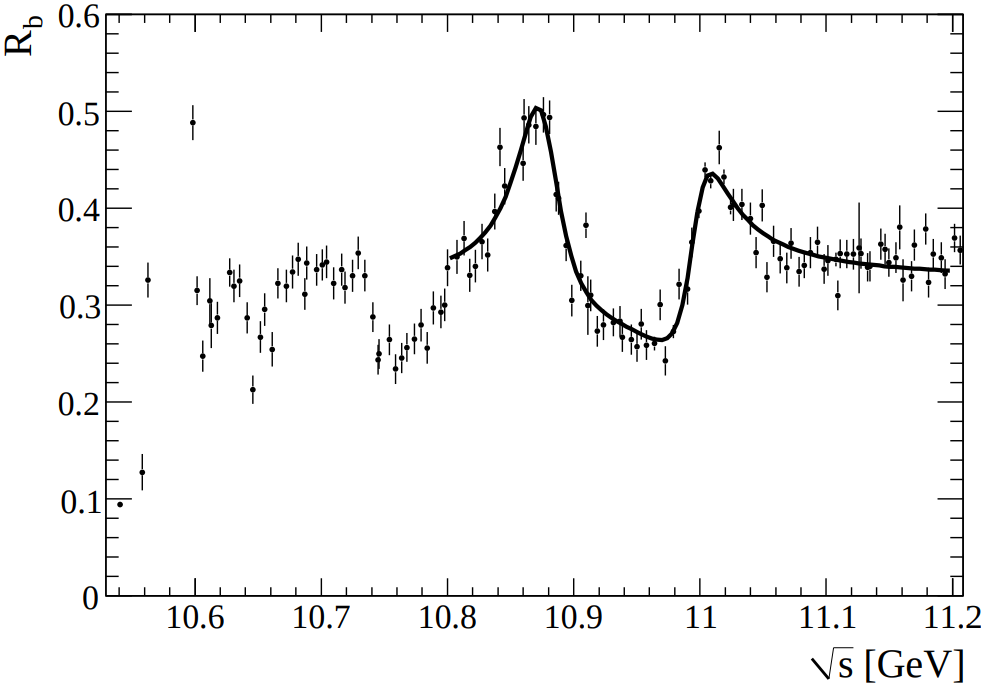"}}
\put(60,0){\includegraphics*[height=37mm]{"./plots/Y10753_BelleII.png"}}
}{
\put(0,0){\includegraphics*[height=45mm]{"./plots/Rb_BaBar.png"}}
\put(70,0){\includegraphics*[height=45mm]{"./plots/Y10753_BelleII.png"}}
}
\ifthenelse{\boolean{captionmode}}{}{
\put(25,38){\footnotesize $\ee\to{hadrons}$ cross-section measured by BaBar~\cite{BaBar:2008cmq} }
\put(100,38){\footnotesize $\ee\to\Upsilon(2S)\pip\pim$ cross-section measured by Belle~\cite{Belle:2019cbt} (right) }
}
\end{picture}
\ifthenelse{\boolean{captionmode}}{
\midplotoffsetcaption
\caption{$\Upsilon(10860)$ and $\Upsilon(11020)$ peaks in the $\ee\to{hadrons}$ cross-section measured by BaBar~\cite{BaBar:2008cmq} (left) and in the $\ee\to\Upsilon(2S)\pip\pim$ cross-section together with the $\Upsilon(10753)$ measured by Belle and \BelleTwo~\cite{Belle-II:2024mjm} (right).
}}{}
\midplotoffsetbottom
\end{figure}
\fi

} 

\cardnotes{
The $\Upsilon(10860)$ and $\Upsilon(11020)$ were first observed as structures in the cross section of \ee annihilation to hadrons by CUSB~\cite{Lovelock:1985nb} and CLEO~\cite{CLEO:1984vfn} already in 1985. 
Much later, both of them were also observed in exclusive decays to $\pi\pi\Upsilon(nS)$ ($n=1,2,3$) and $\pi\pi h_{\b}(nP)$ ($n=1,2)$~\cite{Belle:2019cbt,Belle:2015tbu}. 
In the measurement of the $\ee\to \pip\pim \Upsilon(nS)$ ($n=1,2,3$) cross section, Belle also reported an enhancement, labelled $\Upsilon(10753)$, next to the $\Upsilon(10860)$~\cite{Belle:2019cbt}, which was later confirmed by a more precise measurement of \BelleTwo~\cite{Belle-II:2024mjm}. 
Further evidence for $\Upsilon(10753)$ was gained in the $\ee\to\omega\chi_{b1,2}(1P)$ process by \BelleTwo~\cite{Belle-II:2022xdi} indicating that an earlier hint of the $\Upsilon(10860)\to\omega\chi_{b1,2}(1P)$ decay~\cite{Belle:2014sys} may actually be due to the tail of the $\Upsilon(10753)$.
We refrain from listing decay modes for which resonant production has not been clearly demonstrated.
Branching fractions reported by the PDG are ratios of exclusive cross sections to the inclusive one, which only corresponds to the branching fraction for a narrow, isolated resonance in the absence of any non-resonant production. We now know that this assumption does not hold.
A recent coupled-channel analysis shows that large fractions of $\Upsilon(10860)$ and $\Upsilon(11020)$ decays are so far unobserved, with $B^\ast \bar{B}^{(\ast)}\pi$ being likely candidates~\cite{Husken:2022yik}.

For conventional bottomonium states, it is expected that above the open-bottom threshold the OZI-allowed decays to $B\bar{B}$-type final states far exceed OZI-suppressed decays to hidden-bottom states, which is clearly the case for the $\Upsilon(4S)$. In contrast, the $\Upsilon(10753)$, $\Upsilon(10860)$ and $\Upsilon(11020)$ are the dominant features of the $e^+e^-\to \pi\pi \Upsilon(nS)$ cross sections, shedding some doubt on the interpretation as conventional $\Upsilon$ states. 
However, in the literature in particular the $\Upsilon(10860)$ and $\Upsilon(11020)$ are commonly identified as the conventional $\Upsilon(5S)$ and $\Upsilon(6S)$ states. A bottomonium spin-triplet $\Upsilon(nD)$ assignment of the $\Upsilon(10753)$ is not ruled out by current data~\cite{Li:2019qsg,Husken:2022yik}. A tetraquark assignment is also being discussed~\cite{Wang:2019veq}. In addition, there exist multiple predictions for a bottomonium-hybrid in this mass range, and all three, the $\Upsilon(10753)$, $\Upsilon(10860)$ and $\Upsilon(11020)$ have been hypothesized to be candidates for such an assignment~\cite{Oncala:2017hop,Farina:2020slb,Ryan:2020iog,TarrusCastella:2021pld}.
}
\cardend

%% file: card_Zb.tex
\vtop{ 

\cardstart

\cardtitle{$T_{\b\bbar}$ states}{Meson-like/Hidden Bottom/Isovector}

\ifx\reallybrief 1
\begin{wrapfigure}[0]{r}{0.4\textwidth}
    \vspace{-14mm}
    \includegraphics[width=0.4\textwidth]{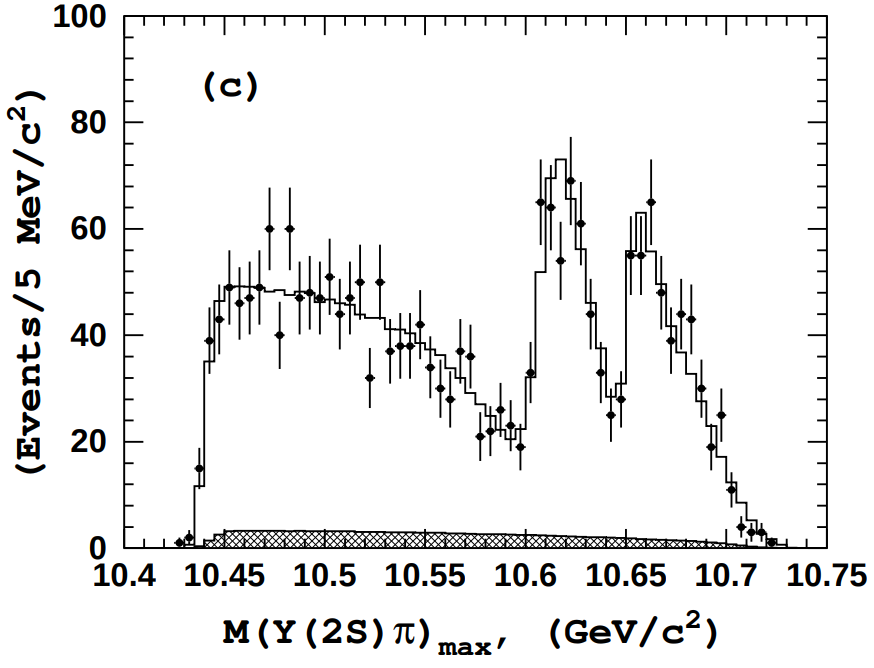}
\wrapcaptionoffset
\caption{The $T_{\b\bbar}^+$ peaks seen in the $\Upsilon(2S)\pip$ by Belle~\cite{Belle:2011aa}.}
\end{wrapfigure}
\fi

\cardproperty{states}{
$\est{T_{\b\bbar1}(10610)^+}$, $\est{T_{\b\bbar1}(10650)^+}$
    \\
    \cardnamealternative{\aka $Z_{b}(10610)^+$, ...
    or $T^{b}_{\Upsilon1}$, ... or $X(10610)$, ...}
}

\cardproperty{quantum numbers}{$I^G(J^{PC}) = 1^+(1^{+-})$}

\cardproperty{minimal quark content}{$[b\bar{b}u\bar{d}]$}

\cardproperty{experiments}{{Belle}}

\cardproperty{production}{$\ee\to T_{\b\bbar}^+\pim$ around the $\Upsilon(10860)$ and $\Upsilon(11020)$
}

\cardproperty{decay modes}{$\pi \Upsilon(nS)$ ($n=1,2,3$), $\pi h_b(nP)$ ($n=1,2)$, 
\\ $B^\ast \bar{B}$ ($T_{\b\bbar1}(10610)$), $B^\ast \bar{B}^\ast$ ($T_{\b\bbar1}(10650)$)}

\cardproperty{nearby thresholds}{$B^{\ast}\bar{B}$, $B^{\ast}\bar{B}^\ast$}

\cardproperty{characteristic widths}{11.5-18.4\mev}

\ifx\reallybrief 1
\else
\ifthenelse{\boolean{formpla}}{
\def\pictW{100}
\def\pictH{40}
}{
\def\pictW{120}
\def\pictH{45}
}

\begin{figure}[H]
\centering
\setlength{\unitlength}{1mm}
\midplotoffsettop
\begin{picture}(\pictW,\pictH)
\ifthenelse{\boolean{formpla}}{
\put(0,0){\includegraphics*[height=40mm]{"./plots/Zb_Y2Spi_highres.png"}}
\put(60,0){\includegraphics*[height=40mm]{"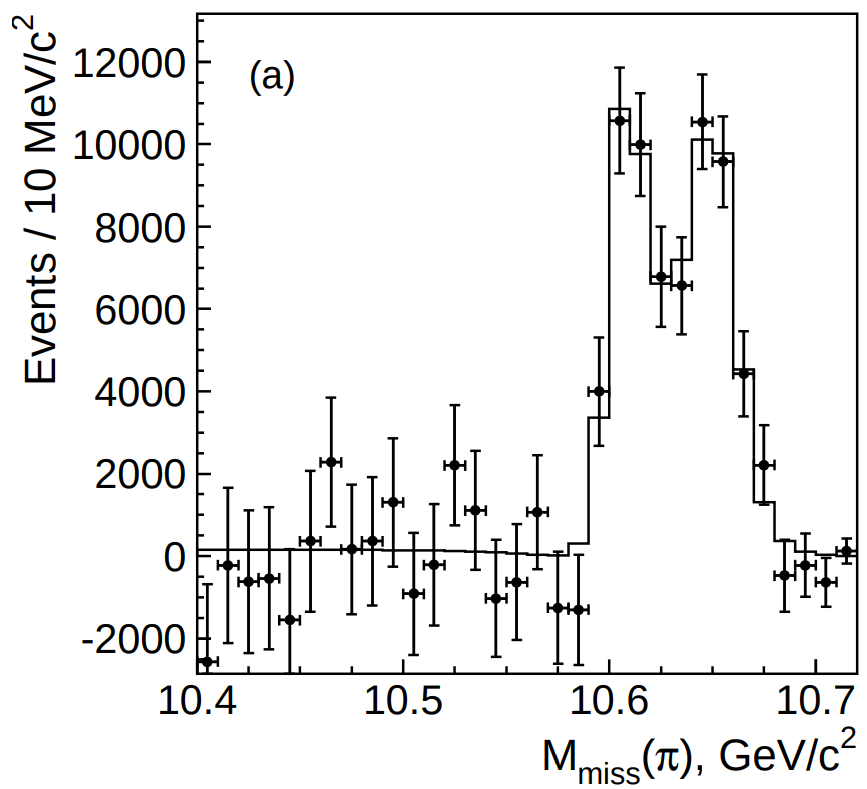"}}
}{
\put(0,0){\includegraphics*[height=45mm]{"./plots/Zb_Y2Spi_highres.png"}}
\put(70,0){\includegraphics*[height=45mm]{"./plots/Zb_hbpi_highres.png"}}
}
\ifthenelse{\boolean{captionmode}}{}{
\put(12,42){\footnotesize $T_{\b\bbar}^+$ in $\Upsilon(2S)\pip$~\cite{Belle:2011aa}}
\put(72,42){\footnotesize $T_{\b\bbar}^+$ in $h_{b}(1P)\pip$~\cite{Belle:2011aa}}
}
\end{picture}
\ifthenelse{\boolean{captionmode}}{
\midplotoffsetcaption
\caption{The $T_{\b\bbar}^+$ peaks seen in the $\Upsilon(2S)\pip$ (left) and $h_{b}(1P)\pip$ (right) modes by Belle~\cite{Belle:2011aa}.
}}{}
\midplotoffsetbottom
\end{figure}
\fi

} 

\cardnotes{
The $T_{\b\bbar}(10610)$ and $T_{\b\bbar}(10650)$ were discovered by the Belle experiment in the $\ee\to\Upsilon(nS)\pip\pim$ and $\ee\to h_b(nP)\pip\pim$ processes~\cite{Belle:2011aa}.
The neutral partner state of the $T_{\b\bbar}(10610)$ was found shortly after, also by Belle~\cite{Belle:2013urd}. A spin-parity of $J^P = 1^+$ is strongly favored from a partial wave analysis of $e^+e^-\to \Upsilon(nS)\pi^+\pi^-$ at 10.866 GeV~\cite{Belle:2014vzn}.
Similar to the $T_{c\bar{c}1}(3900)^+$ and $T_{c\bar{c}}(4020)^+$ at the $D^\ast \bar{D}$ and $D^\ast \bar{D}^\ast$ thresholds, the $T_{b\bar{b}1}(10610)$ and its heavier cousin, the $T_{b\bar{b}1}(10650)$, sit right at the $B^\ast \bar{B}$ and $B^\ast \bar{B}^\ast$ thresholds and decay dominantly to those channels~\cite{Belle:2015upu}. As such, they are widely viewed as strongly related to the corresponding $T_{\c\cbar}$ states in the charmonium-sector, likely sharing a common explanation. The first observation actually pre-dates the discovery of their charmonium look-a-likes. In another striking similarity, the $T_{b\bar{b}1}(10610)$ and $T_{b\bar{b}1}(10650)$ appear to be correlated to the the $\Upsilon(10860)$ and $\Upsilon(11020)$ peaks in the exclusive $\ee \to \Upsilon(nS)\pi\pi$ and $e^+e^-\to h_b(nP)\pi\pi$ cross-sections~\cite{Belle:2015tbu}, 
just like the $T_{c\bar{c}1}(3900)$ and $T_{c\bar{c}1}(4020)$ are correlated to the $\psi(4230)$ and $\psi(4360)$. However, the $\Upsilon(10860)$ and $\Upsilon(11020)$ are in contrast to the charmonium-like states more commonly interpreted as the conventional $\Upsilon(5S)$ and $\Upsilon(6S)$ bottomonium excitations.

In the literature, a molecular interpretation of the $T_{\b\bbar}$ states is favored~\cite{Bondar:2011ev,Cleven:2013sq,Dias:2014pva,Goerke:2017svb,Ding:2020dio}. It is however argued that such an interpretation requires spin-partner states that have yet to be found in the experiment~\cite{Voloshin:2011qa,Mehen:2011yh,Baru:2017gwo}. 
Meanwhile, an interpretation as compact tetraquark states has not been ruled out~\cite{Ali:2014dva,Agaev:2017lmc}. 
In Ref.~\cite{Kang:2016ezb}, the compositeness of the $T_{\b\bbar}$ states was studied, 
indicating strong molecular component.
Only few lattice-QCD studies relating to the $T_{\b\bbar}$ exist, finding an attractive $B^\ast \bar B$ interaction~\cite{Peters:2016wjm,Prelovsek:2019ywc,Sadl:2021bme}, that might support a molecular interpretation.
}

\cardend

%% file: card_JpsiJpsi.tex
\vtop{ 

\cardstart

\cardtitle{$T_{\c\cbar\c\cbar}$ states}{Meson-like/Hidden Double Charm}

\ifx\reallybrief 1
\begin{wrapfigure}[0]{r}{0.4\textwidth}
    \vspace{-14mm}
    \includegraphics[width=0.4\textwidth]{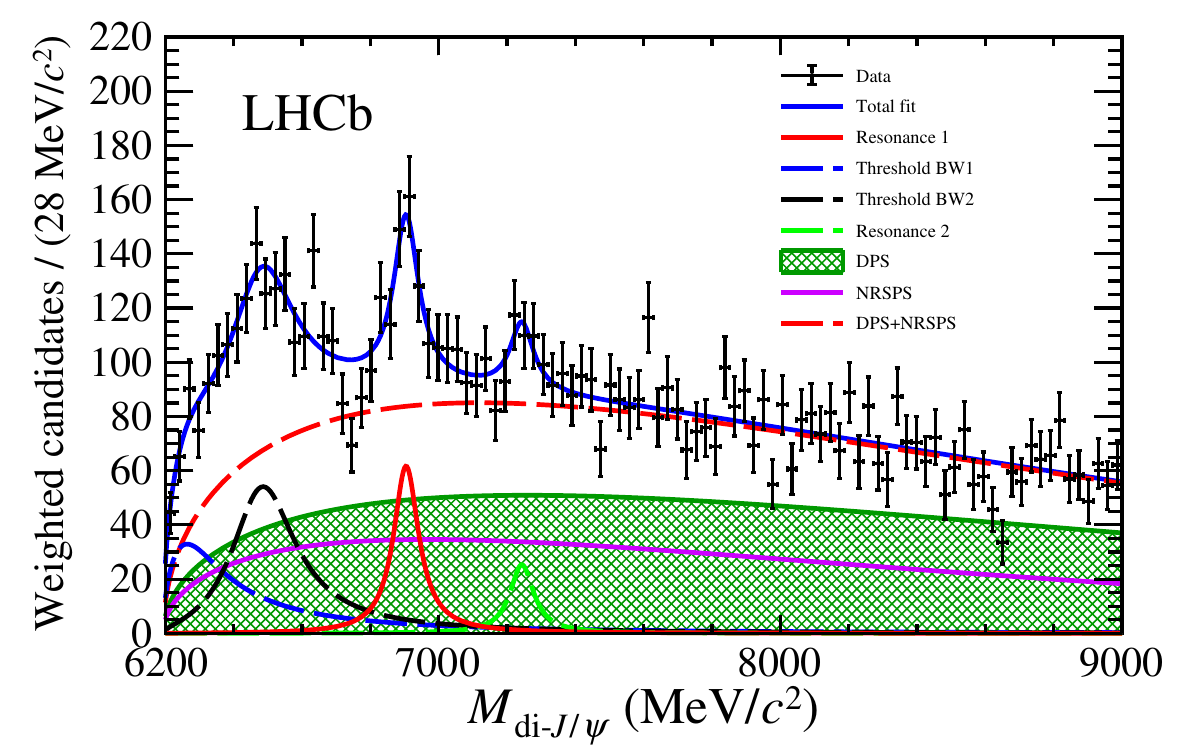}
\wrapcaptionoffset
\caption{Discovery of the structures in $\jpsi\jpsi$ by LHCb~\cite{LHCb:2020bwg}.}
\end{wrapfigure}
\fi

\cardproperty{states}{
    $T_{\c\cbar\c\cbar}(6550)^0$, $\est{T_{\c\cbar\c\cbar}(6900)^0}$, $T_{\c\cbar\c\cbar}(7290)^0$
    \\
    \cardnamealternative{\aka $T_{\psi\psi}(6900)^0$, ... or $X(6900)$, ...}
}

\cardproperty{quantum numbers}{$I^G(J^{PC}) = 0^+(?^{?+})$}

\cardproperty{minimal quark content}{$[\c\cbar\c\cbar]$}

\cardproperty{experiments}{LHCb, ATLAS, CMS}

\cardproperty{production}{prompt $pp$ collisions}

\cardproperty{decay modes}{$\jpsi\jpsi$}

\cardproperty{nearby thresholds}{$\jpsi\psitwos$, $\chi_{\c0}\chi_{\c0}$, ... }

\cardproperty{characteristic widths}{80-191\mev}

\ifx\reallybrief 1
\else
\ifthenelse{\boolean{formpla}}{
\def\pictW{100}
\def\pictH{30}
}{
\def\pictW{125}
\def\pictH{38}
}

\begin{figure}[H]
\centering
\setlength{\unitlength}{1mm}
\midplotoffsettop
\begin{picture}(\pictW,\pictH)
\ifthenelse{\boolean{formpla}}{
\put(0,0){\includegraphics*[height=30mm]{"./plots/JpsiJpsi_LHCb.pdf"}}
\put(60,0){\includegraphics*[height=30mm]{"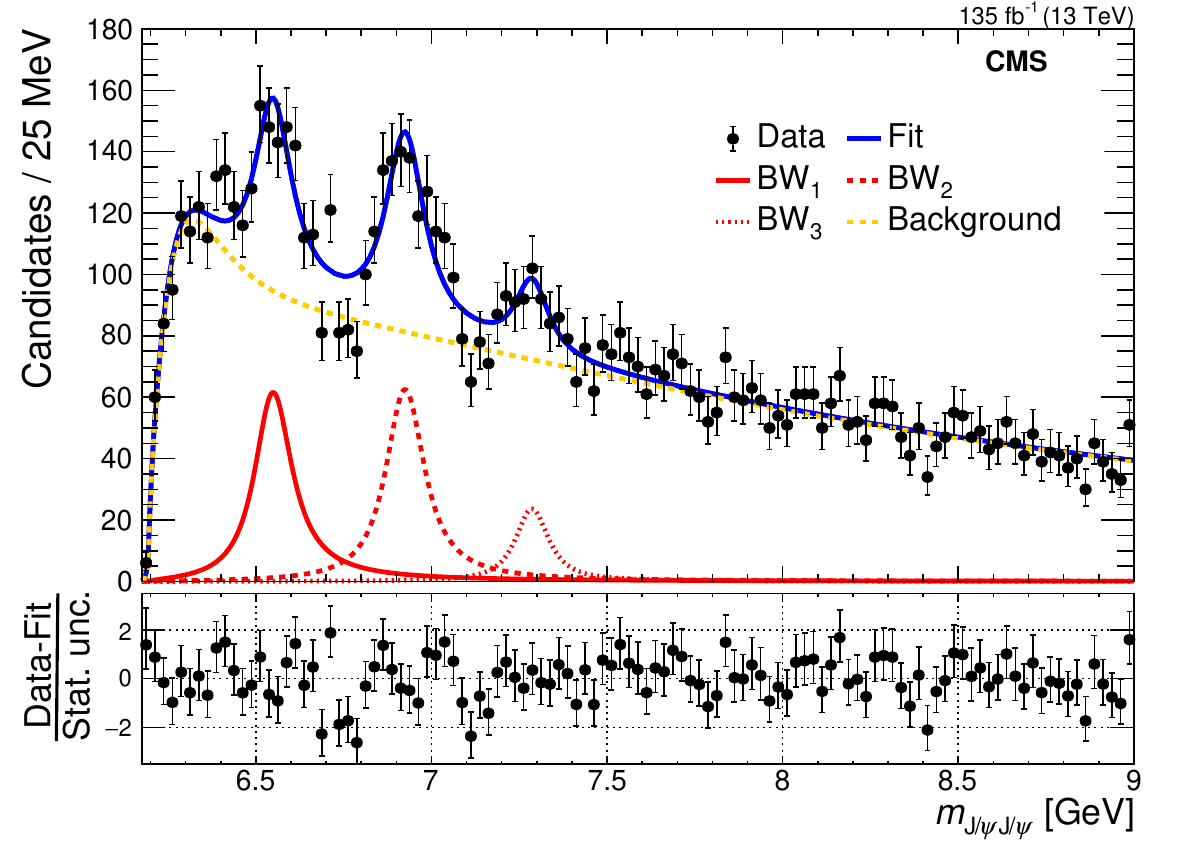"}}
}{
\put(0,0){\includegraphics*[height=38mm]{"./plots/JpsiJpsi_LHCb.pdf"}}
\put(70,2){\includegraphics*[height=36mm]{"./plots/JpsiJpsi_CMS.pdf"}}
}%
\ifthenelse{\boolean{captionmode}}{}{
\put(15,31){\footnotesize in LHCb~\cite{LHCb:2020bwg}}
\put(75,31){\footnotesize in CMS~\cite{CMS:2023owd}}
}
\end{picture}
\ifthenelse{\boolean{captionmode}}{
\midplotoffsetcaption
\caption{$\jpsi\jpsi$ mass spectra from $\pp$ collisions in LHCb~\cite{LHCb:2020bwg} (left) and CMS~\cite{CMS:2023owd} (right).
}}{}
\midplotoffsetbottom
\end{figure}
\fi

} 

\cardnotes{
Resonant structures in $\jpsi\jpsi$ were first observed in 2020 by LHCb~\cite{LHCb:2020bwg}, with the most prominent peak at 6.9\gev and hints of structures near 6.5\gev and~7.2\gev. Later, CMS~\cite{CMS:2023owd} and ATLAS~\cite{ATLAS:2023bft} independently confirmed the existence of the resonance at~6.9\gev. 
CMS also claimed observation of a second state around 6.6\gev with a significance larger than 6$\sigma$ and an evidence for a third state around 7.29\gev at the level of 4$\sigma$.
ATLAS have also found hints of the same states in the $\psitwos\jpsi$ channel. 

The measured yields, masses and widths of the $T_{\c\cbar\c\cbar}$ candidates were found to strongly depend on the treatment of interference between the resonances in one-dimensional fits to the invariant mass distributions. In addition, the resonance picture might appear to be much more complicated with many resonances expected by 
different theoretical models~\cite{Faustov:2020qfm,Giron:2020wpx,Jin:2020jfc,Lu:2020cns}. Therefore the experimental values, especially for the resonances near 6.6 and 7.3\gev, should be taken with caution until additional measurements in complementary decay channels provide a cleaner picture.

A number of theoretical calculations explaining the resonances as compact tetraquarks, molecular states or other more exotic states are available in the literature~\cite{Wan:2020fsk,Jin:2020jfc,Lu:2020cns,Dong:2020nwy,Wang:2020wrp}, with some works predicting the existence of such states already in 1975~\cite{Iwasaki} and 1982~\cite{PhysRevD.25.2370}.
A more complete list of related works can be found in Ref.~\cite{PDG2023_Non_qq_review}.}

\cardend

%% file: card_Dstar.tex
\vtop{ 

\cardstart

\cardtitle{$\D_{s0/1}^{*}$ states}{Meson-like/Open Single Charm}

\ifx\reallybrief 1
\begin{wrapfigure}[0]{r}{0.4\textwidth}
    \vspace{-14mm}
    \includegraphics[width=0.4\textwidth]{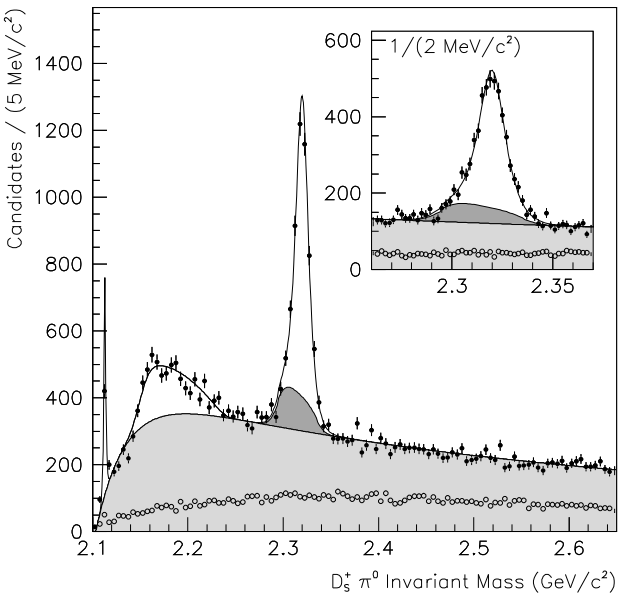}
\wrapcaptionoffset
\caption{$D^{*}_{\s0}(2317)^+\to\Ds\piz$ signal at BaBar~\cite{BaBar:2006eep}.}
\end{wrapfigure}
\fi

\cardproperty{states}{
    \cardnamelist{
    \item $I(J^P)=0(0^{+})$: $\est{\D_{\s0}^{*}(2317)^+}$
    \item $I(J^P)=0(1^{+})$: $\est{\D_{\s1}(2460)^+}$
    }
}

\cardproperty{minimal quark content}{$[\c\sbar]$, or possibly $[\c\sbar\q\qbar]$}

\cardproperty{experiments}{BaBar, CLEO, Belle, BESIII}

\cardproperty{production}{$\ee\to\D_{\s J} X$, $\B\to\D_{\s J}\Db$}

\cardproperty{decay modes}{$\Ds\piz$ for $\D_{\s0}^{*}(2317)$,
\\ $\Dstars\piz$, $\Ds\gamma$, $\Ds\pip\pim$ for $\D_{\s1}(2460)$
}

\cardproperty{nearby thresholds}{$\D\kaon$, $\D\Kstar$}

\cardproperty{characteristic widths}{$<3.8\mev$}

\ifthenelse{\boolean{formpla}}{
\def\pictW{100}
\def\pictH{40}
}{
\def\pictW{120}
\def\pictH{50}
}

\ifx\reallybrief 1
\else
\begin{figure}[H]
\centering
\setlength{\unitlength}{1mm}
\midplotoffsettop
\begin{picture}(\pictW,\pictH)
\ifthenelse{\boolean{formpla}}{
\put(0,0){\includegraphics*[height=40mm]{"./plots/Dspi0_BaBar.png"}}
\put(60,0){\includegraphics*[height=40mm]{"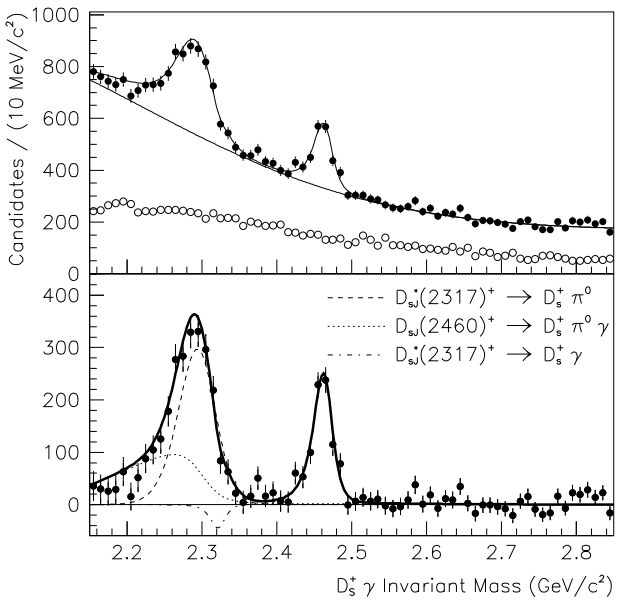"}}
}{
\put(0,0){\includegraphics*[height=50mm]{"./plots/Dspi0_BaBar.png"}}
\put(70,0){\includegraphics*[height=50mm]{"./plots/Dsgamma_BaBar.png"}}
}
\ifthenelse{\boolean{captionmode}}{}{
\put(2,42){\footnotesize $D^{*}_{\s0}(2317)^+\to\Ds\piz$ at BaBar~\cite{BaBar:2006eep}}
\put(62,42){\footnotesize $D_{\s1}(2460)^+\to\Ds\gamma$ at BaBar~\cite{BaBar:2006eep}}
}
\end{picture}
\ifthenelse{\boolean{captionmode}}{
\midplotoffsetcaption
\caption{$D^{*}_{\s0}(2317)^+\to\Ds\piz$ (left) and $D_{\s1}(2460)^+\to\Ds\gamma$ (right) signals at BaBar~\cite{BaBar:2006eep}.
}}{}
\midplotoffsetbottom
\end{figure}
\fi

} 

\cardnotes{
The two states $D^\ast_{s0}(2317)^+$ and $D^\ast_{s1}(2460)^+$ were first observed by the BaBar~\cite{BaBar:2003oey,BaBar:2003cdx} and CLEO~\cite{CLEO:2003ggt} experiments. Later, they were confirmed by Belle~\cite{Belle:2003guh} and BESIII~\cite{BESIII:2017vdm}.
Both have positive parity and in principle could correspond to the conventional $P-$wave excitations of the regular \Ds $[\c\sbar]$ mesons. However, the $D^\ast_{s0}(2317)^+$ and $D^\ast_{s1}(2460)^+$ appear to have masses lower than expectations for excited $D_s$ mesons by 50-200\mev (see~\cite{Godfrey:2003kg} for further references) and also appear to be too narrow, with widths below 3.7 MeV. Therefore, exotic explanations as compact tetraquark states~\cite{Cheng:2003kg,Maiani:2004vq} or $\D\kaon^{(*)}$ molecules~\cite{Barnes:2003dj} have been proposed. At the same time, a conventional interpretation is not excluded as the masses and widths could be affected by nearby $\D\kaon^{(*)}$ thresholds via cusp effects~\cite{Godfrey:2003kg}. 
The small widths can be explained by the fact that possible strong decays are isospin-violating and hence suppressed.
}

\cardend

%% file: card_Tcs.tex
\vtop{ 

\cardstart

\cardtitle{$T_{\c\s/\c\sbar}$ states}{Meson-like/Open Single Charm}

\ifx\reallybrief 1
\begin{wrapfigure}[0]{r}{0.4\textwidth}
    \vspace{-14mm}
    \includegraphics[width=0.4\textwidth]{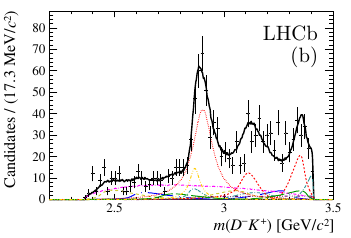}
\wrapcaptionoffset
\caption{$T_{\c\s0,1}(2900)^0$ in ${\Bm\to\Dp\Dm\Km}$~\cite{LHCb:2020pxc}.}
\end{wrapfigure}
\fi

\cardproperty{states}{
    \cardnamelist{
    \item $I(J^P)=?(0^{+})$: $T_{\c\s0}(2900)^0$, $T_{\c\sbar0}(2900)^0$, $T_{\c\sbar0}(2900)^{++}$
    
    \item $I(J^P)=?(1^{-})$: $T_{\c\s1}(2900)^0$
    }
    \cardnamealternative{\aka $T_{\c\s0}^{*}(2870)^0$,  $T_{\c\s1}^{*}(2900)^0$, ...
    or $X_0(2900)$, $X_1(2900)$, ...}
}

\cardproperty{minimal quark content}{$[\c\s\qbar\qbar']$, $[\c\sbar\q\qbar']$}

\cardproperty{experiments}{LHCb}

\cardproperty{production}{$\Bm\to\D^{(\star)+}\Dm\Km$,  $\Bd\to\Dzb\Ds\pim$,  $\Bp\to\Dm\Ds\pip$}

\cardproperty{decay modes}{$\Dp\Km$, $\Ds\pim$, $\Ds\pip$}

\cardproperty{nearby thresholds}{$\Dstar\Kstar$, $\D_1\kaon$
}

\cardproperty{characteristic widths}{57-136\mev}

\ifx\reallybrief 1
\else
\ifthenelse{\boolean{formpla}}{
\def\pictW{100}
\def\pictH{30}
}{
\def\pictW{120}
\def\pictH{38}
}

\begin{figure}[H]
\centering
\setlength{\unitlength}{1mm}
\midplotoffsettop
\begin{picture}(\pictW,\pictH)
\ifthenelse{\boolean{formpla}}{
\put(0,0){\includegraphics*[height=31mm]{"./plots/Tcs2900_DK.pdf"}}
\put(60,0){\includegraphics*[height=30mm]{"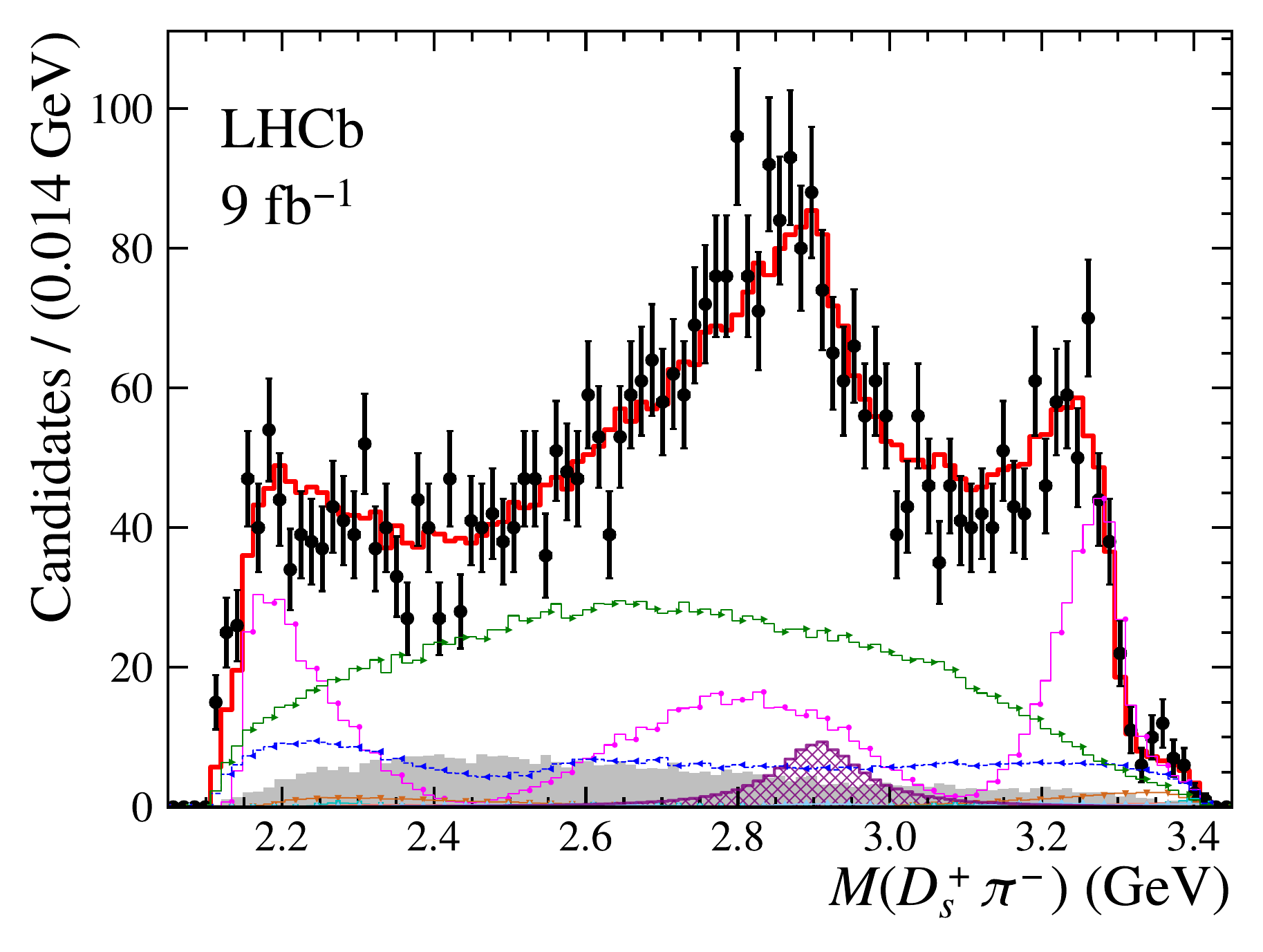"}}
}{
\put(0,0){\includegraphics*[height=38mm]{"./plots/Tcs2900_DK.pdf"}}
\put(70,0){\includegraphics*[height=37mm]{"./plots/Tcs2900_Dspim.pdf"}}
}
\ifthenelse{\boolean{captionmode}}{}{
\put(2,32){\footnotesize $T_{\c\s0,1}(2900)^0$ in $\Bm\to\Dp\Dm\Km$~\cite{LHCb:2020pxc}}
\put(62,32){\footnotesize $T_{\c\sbar0}(2900)^0$ in $\Bd\to\Dzb\Ds\pim$~\cite{LHCb:2022sfr}}
}

\end{picture}
\ifthenelse{\boolean{captionmode}}{
\midplotoffsetcaption
\caption{$T_{\c\s0,1}(2900)^0$ in $\Bm\to\Dp\Dm\Km$~\cite{LHCb:2020pxc} (left) and $T_{\c\sbar0}(2900)^0$ in $\Bd\to\Dzb\Ds\pim$~\cite{LHCb:2022sfr} (right) signals seen by LHCb.
}}{}
\midplotoffsetbottom
\end{figure}

\fi

} 

\cardnotes{

The observation and spin-parity assignments of the flavor-exotic $T_{cs}$ and $T_{c\sbar}$ states are based on amplitude analyses of the ${\B\to\Db\D\kaon}$ and ${\B\to\Db\D_{\s}\pion}$ decays~\cite{LHCb:2020bls,LHCb:2020pxc,LHCb:2022sfr,LHCb:2022lzp,LHCb:2024vfz}.
A resonant nature of these states was confirmed
from the Argand diagram, showing the characteristic phase-motion that is
expected for a resonance.
\\
Because of the decay of the $T_{\c\s0,1}(2900)^0$ states to $\Dp\Km$,
their minimal quark content is $[\c\s\ubar\dbar]$. Similarly, the $T_{\c\sbar0}(2900)^{++}$ decay to $\Ds\pip$ clearly signals a minimal quark content of $[\c\sbar\u\dbar]$, while the $T_{\c\sbar0}(2900)^{0}$ decay to $\Ds\pim$ corresponds to a $[\c\sbar\d\ubar]$ content. Thus, all four are manifestly exotic states with open charm.

The two $T_{\c\sbar0}(2900)^{0/++}$ states can be considered as $SU(3)_{F}$ partners of the $T_{cs0}(2900)^0$ state. 
However, the subtle difference in masses and significant difference in widths may point to a non-trivial nature of such a relationship~\cite{Dmitrasinovic:2023eei}.
Among interpretations for these $T_{\c\s}$ and $T_{\c\sbar}$ states are compact tetraquarks~\cite{Karliner:2020vsi,Guo:2021mja}, 
$\D^{*}K^{*}$ and $\D_{1}K$ molecules~\cite{Molina:2020hde} 
and threshold cusps or triangle singularities~\cite{Liu:2020orv,Burns:2020xne}.
}

\cardend

%% file: card_Tcc.tex
\vtop{ 

\cardstart

\cardtitle{The $T_{\c\c}(3875)^+$}{Meson-like/Open Double Charm}

\begin{wrapfigure}[0]{r}{0.4\textwidth}
    \vspace{-14mm}
\ifthenelse{\boolean{captionmode}}{}{
    {\color{black} \footnotesize \phantom{000} Discovery of the \Tcc~\cite{LHCb-PAPER-2021-031}} \\
}
    \includegraphics[width=0.4\textwidth]{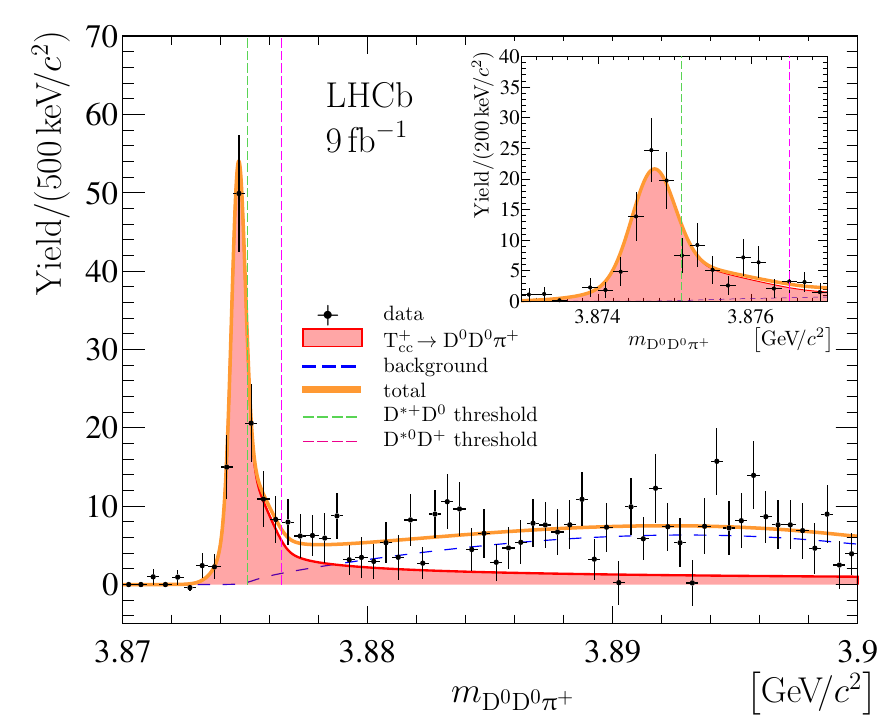}
\ifthenelse{\boolean{captionmode}}{
\wrapcaptionoffset
\caption{Discovery of the \Tcc by LHCb~\cite{LHCb-PAPER-2021-031}.}
}{}
\end{wrapfigure}

\cardproperty{quantum numbers}{$I(J^P) = ?(?^?)$, likely $0(1^+)$}

\cardproperty{minimal quark content}{$[\c\c\ubar\dbar]$}

\cardproperty{experiments}{LHCb}

\cardproperty{production}{prompt \pp collisions}

\cardproperty{decay modes}{ \\ $\Dz\Dz\pip$ and $\Dz\Dp(\piz/\g)$ via $\D\Dstar$}

\cardproperty{nearby thresholds}{$\Dz\Dstarp$, $\Dp\Dstarz$}

\cardproperty{characteristic width}{${\sim}50\kev$}

} 

\ifthenelse{\boolean{formpla}}{
\vspace{2mm}
}{
\vspace{12mm}
}

\ifthenelse{\boolean{captionmode}}
{\vspace{2mm}}
{\vspace{2mm}}

\cardnotes{
In what is likely the most exciting discovery in quarkonium spectroscopy since the $\chi_{c1}(3872)$,
a narrow structure, the $T_{\c\c}(3875)^+$, is observed by LHCb in the invariant mass of the $\Dz\Dz\pip$ system produced directly in \pp collisions~\cite{LHCb-PAPER-2021-031}. 
Due to the proximity of the state to the $\Dz\Dstarp$ threshold, a fit to the $\Dz\Dz\pip$ mass distribution was performed with a specialized Breit-Wigner model, taking into account threshold effects and ensuring both unitarity and analyticity~\cite{LHCb-PAPER-2021-032}.
The mass of the state relative to the $\Dz\Dstarp$ threshold was measured to be 
${\delta m = -359 \pm 40 ^{+9}_{-6}\kev}$.
For this value of the mass, the model calculation yields a width of only $48 \pm 2 ^{+0} _{-14} \kev$, which is $10-1000$ times more narrow than most other exotic hadrons. 
The aforementioned model relies on two assumptions: that the state has isospin and spin-parity $(I)J^P=(0)1^+$, in accordance with the expectation for a $\c\c\ubar\dbar$ ground state of either compact or molecular nature;
and that it decays strongly via $\Dz\Dstarp$ and $\Dp\Dstarz$ with the \Dstar being off-shell.
These assumptions are supported by a simultaneous successful description of the $\Dz\Dz\pip$ and $\Dz\pip$ mass distributions in the $\Tcc\to\Dz\Dz\pip$ decay, as well as of the $\Dz\Dz$ and $\Dz\Dp$ mass distributions
(reflections from decays to $\Dz\Dz\pip$ and $\Dz\Dp\piz(\g)$, where either a pion or photon is not reconstructed). 
Support for the isospin-0 assignment is provided by the non-observation of any peaking structures in the $\Dp\Dp$ and $\Dp\Dz\pip$ systems. 

Altogether, the current information allows to identify the observed structure to very likely be a $[\c\c\ubar\dbar]$ tetraquark ground state with $I(J^P)=0(1^+)$ quantum numbers
confirming theoretical calculations~\cite{Semay:1994ht,Janc:2004qn,Li:2012ss,Liu:2019stu,Karliner:2017qjm,Junnarkar:2018twb} (see more in Ref~\cite{LHCb-PAPER-2021-032}) including the first predictions dating back to 1982~\cite{PhysRevD.25.2370,BALLOT1983449}.
Given the perfect agreement with the model that only considers decays via $\D\Dstar$, it is likely that the \Tcc is predominantly a $\D\Dstar$ molecule.
This is also supported by calculations in a realistic model which allows for both diquark-antidiquark and molecule configurations~\cite{Janc:2004qn}, especially given the small binding energy observed.
Nonetheless, a compact component may also be present and its size is a question for future studies.

The \Tcc holds a special place among all known exotic hadrons.
Its mass is measured with unprecedented precision compared to all known exotic hadrons, thus not only giving a stringent test on various theoretical models, but providing unique input for further fine-tuning. 
Second, in contrast to all other known hadron molecular candidates it cannot decay via quark-antiquark annihilation and hence provides a much cleaner testing-ground for 
exotic hadron studies.
One can therefore claim a similar, or in some sense even better, level of understanding of the \Tcc nature as for the \theX after 20 years of thorough investigations, despite having observed only around $150$ candidate events.
The observation of the \Tcc just below the $\Dz\Dstarp$ threshold suggests, according to a number of models~\cite{Semay:1994ht,Karliner:2017qjm}, that its partner with $\b\c\ubar\dbar$ quark content might lie below threshold for strong and electromagnetic decays and hence be long-lived, i.e. be the long-awaited holy grail of exotic hadron spectroscopy with realistic prospects for discovery in the near future~\cite{TbcInRun3}.

}

\cardend

%% file: card_Pc.tex
\vtop{ 

\cardstart

\cardtitle{$P_{\c\cbar}$ states
\cardnamealternative{(\aka $P_{\psi}^{N}$ or $P_c$)}
}{Baryon-like/Hidden Charm/Isospin=$\nicefrac{1}{2}(\nicefrac{3}{2})$}

\begin{wrapfigure}[0]{r}{0.4\textwidth}
    \vspace{-13mm}
\ifthenelse{\boolean{captionmode}}{}{
    {\color{black} \footnotesize \phantom{000} $P_{\c\cbar}$ states in  $\Lambda_b \rightarrow \jpsi \proton \Km$~\cite{LHCb-PAPER-2019-014}} \\
}
    \includegraphics[width=0.4\textwidth]{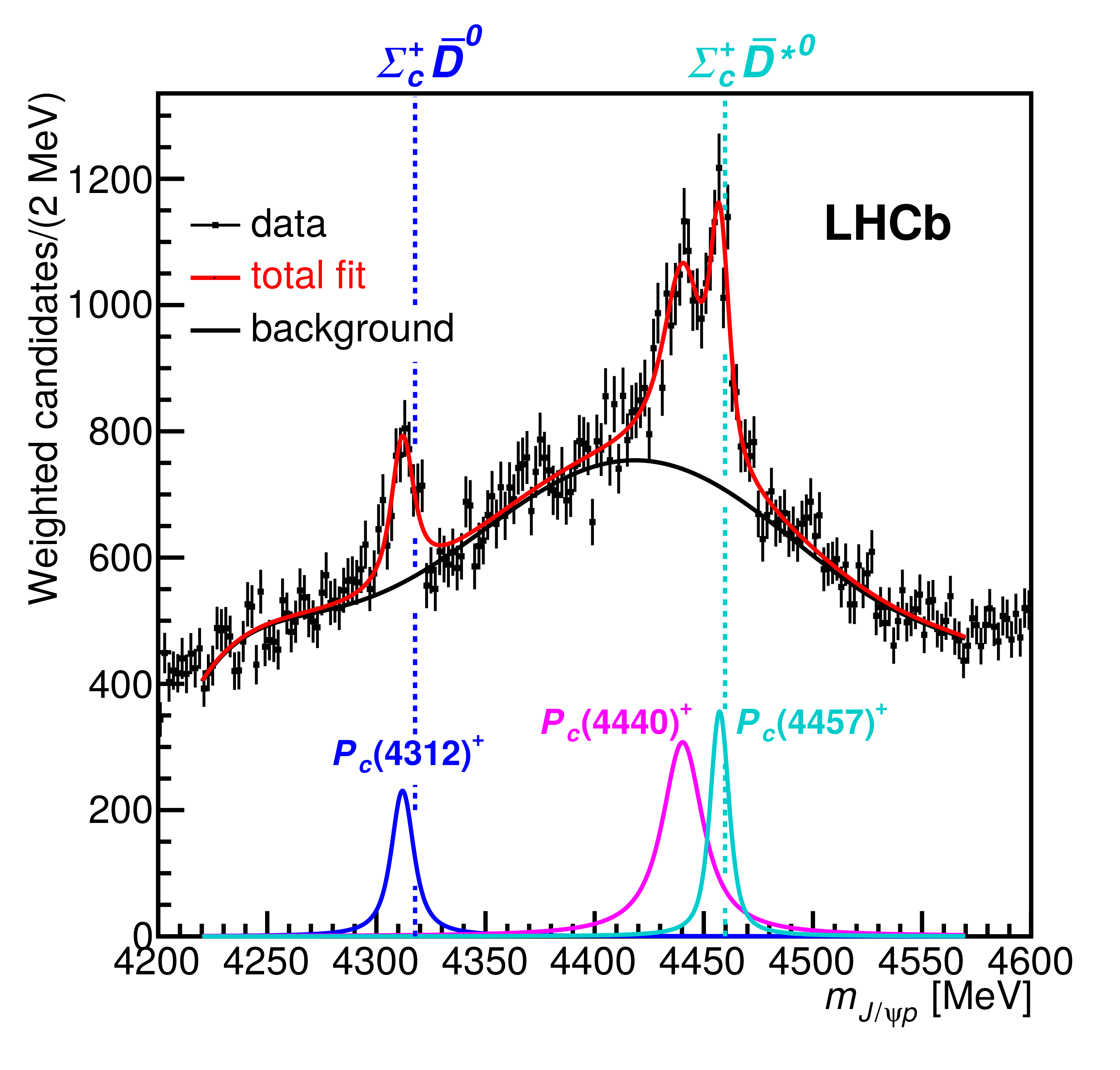}
\ifthenelse{\boolean{captionmode}}{
\wrapcaptionoffset
\caption{$P_{\c\cbar}$ states in $\Lambda_b \rightarrow \jpsi \proton \Km$ at LHCb~\cite{LHCb-PAPER-2019-014}.}
}{}
\end{wrapfigure}

\cardproperty{states}{
 $P_{\c\cbar}(4312)^+$, $P_{\c\cbar}(4440)^+$, $P_{\c\cbar}(4457)^+$, \\ $P_{\c\cbar}(4380)^+$, $P_{\c\cbar}(4337)^+$
    
}

\cardproperty{minimal quark content}{$[\c\cbar\u\u\d]$}

\cardproperty{experiments}{LHCb}

\cardproperty{production}{$\Lambda_b\to\jpsi\proton\Km$, 
\\ likely $\Lambda_b\to\jpsi\proton\pim$ ($P_{\c\cbar}(4440|4457)$),
\\ $\Bs\to\jpsi\proton\bar{\proton}$} ($P_{\c\cbar}(4337)$)

\cardproperty{decay modes}{$\jpsi \proton$}

\cardproperty{nearby threshold}{$\Sigma_c^+ \bar{D}^{(*)0}$}

\cardproperty{characteristic widths}{10-30\mev \\ and ${\sim}205\mev$ ($P_{\c\cbar}(4380)$)}

} 

\ifthenelse{\boolean{formpla}}{
\vspace{0mm}
}{
\vspace{8mm}
}

\cardnotes{
First pentaquark states were observed by LHCb in 2015 in the decay ${\Lambda_b \rightarrow \jpsi \proton \Km}$~\cite{LHCb:2015yax}, indicating a minimal quark content of $\c \cbar \u \u \d$.
A full amplitude analysis
revealed the existence of two states, the $P_{\c\cbar}(4380)^+$ and $P_{\c\cbar}(4450)^+$, with widths of $205 \pm 18 \pm 86$ MeV and $39 \pm 5 \pm 19\mev$. Preferred quantum numbers are $J^P = \frac{3}{2}^-$ and $J^P = \frac{5}{2}^+$, respectively.  However, other spin-parity hypotheses corresponding to combinations of $(3/2^+, 5/2^-)$ and $(5/2^+, 3/2^-)$ are not excluded.
In 2019, with an updated analysis performed with the full 9 fb$^{-1}$ LHCb dataset, another state at a mass of $4312\mev$ was observed~\cite{LHCb-PAPER-2019-014}. In addition, the peak at a mass of $4450\mev$ was resolved into two separate peaks with masses of $4440$ and $4457\mev$, respectively.
The analysis used a one-dimensional fit to the $\jpsi\proton$ invariant mass distribution to determine the masses and widths of the peaking structures. For the determination of the quantum numbers, a full amplitude analysis is required. The presence of the thresholds of the $\Sigma_c^+ \bar{D}^{(*)0}$ systems just a few MeV above the peak mass values could be accidental, but more likely indicates a molecular nature~\cite{Yang:2011wz,Wu:2010jy,Karliner:2015ina,Penta:molecules_PhysRevD.92.094003}. This is also supported by their relatively small widths, that call for some sort of width suppression mechanism. 
Other possible interpretations are compact $\c\cbar\u\u\d$ states~\cite{Penta:compact,Lebed:2015tna,Anisovich:2015cia,Li:2015gta} or triangle cusps~\cite{Penta_triangle:PhysRevD.92.071502}. 
However, LHCb demonstrated the cusp interpretation to be unlikely at least for $P_{\c\cbar}(4312)^+$ and $P_{\c\cbar}(4440)^+$~\cite{LHCb-PAPER-2019-014}.

An additional pentaquark candidate, $P_{\c\cbar}(4337)^+$, is observed in the $\Bs\to J/\psi \proton \bar{\proton}$ decay, again in the $\jpsi \proton$ system, with a mass of $4337\mev$ and a width of $30\mev$~\cite{LHCb-PAPER-2021-018}. 
This state is incompatible with the pentaquark states observed in $\Lambda_b$ decays.  Different theoretical interpretations have been put forward to explain this feature: it could either be a compact pentaquark~\cite{PhysRevD.104.114028, germani2024simple} due to the different internal spin structure of the di-quark pair, or a triangle cusp~\cite{PhysRevD.104.L091503} caused by the nearby $\Sigma_c\bar{D}$ or $\Lambda_c D^*$ thresholds.
A molecular interpretation is unlikely since the mass of the state is $20-40\mev$ above the aforementioned thresholds.
In the compact pentaquark interpretation, the expected $J^P$ would be $1/2^+$, while for the cusp interpretation, one would expect $J^P=1/2^-$ -- these possibilities can be distinguished by LHCb in the future. 
An additional possibility to rule out explanations related to possible triangle singularities is to search for pentaquark candidates via other production mechanisms, for instance via photoproduction $\gamma \proton \to \jpsi \proton$ as it is pursued by the GlueX experiment~\cite{GlueX:2019mkq,Duran:2022xag,GlueX:2023pev,JointPhysicsAnalysisCenter:2023qgg,Strakovsky:2023kqu}.
}

%% file: card_Pcs.tex
\vtop{ 

\cardstart

\cardtitle{$P_{\c\cbar\s}$ states 
\cardnamealternative{(\aka $P^{\Lambda}_{\psi\s}$ or $P_{\c\s}$)}
}{Baryon-like/Hidden Charm/Isospin=0(1)}

\begin{wrapfigure}[0]{r}{0.4\textwidth}
    \vspace{-10mm}    
\ifthenelse{\boolean{captionmode}}{}{
    {\color{black} \footnotesize \phantom{000} Discovery of the $P_{\c\cbar \s}(4338)^0$~\cite{LHCb-PAPER-2022-031}} \\
}
    \includegraphics[width=0.4\textwidth]{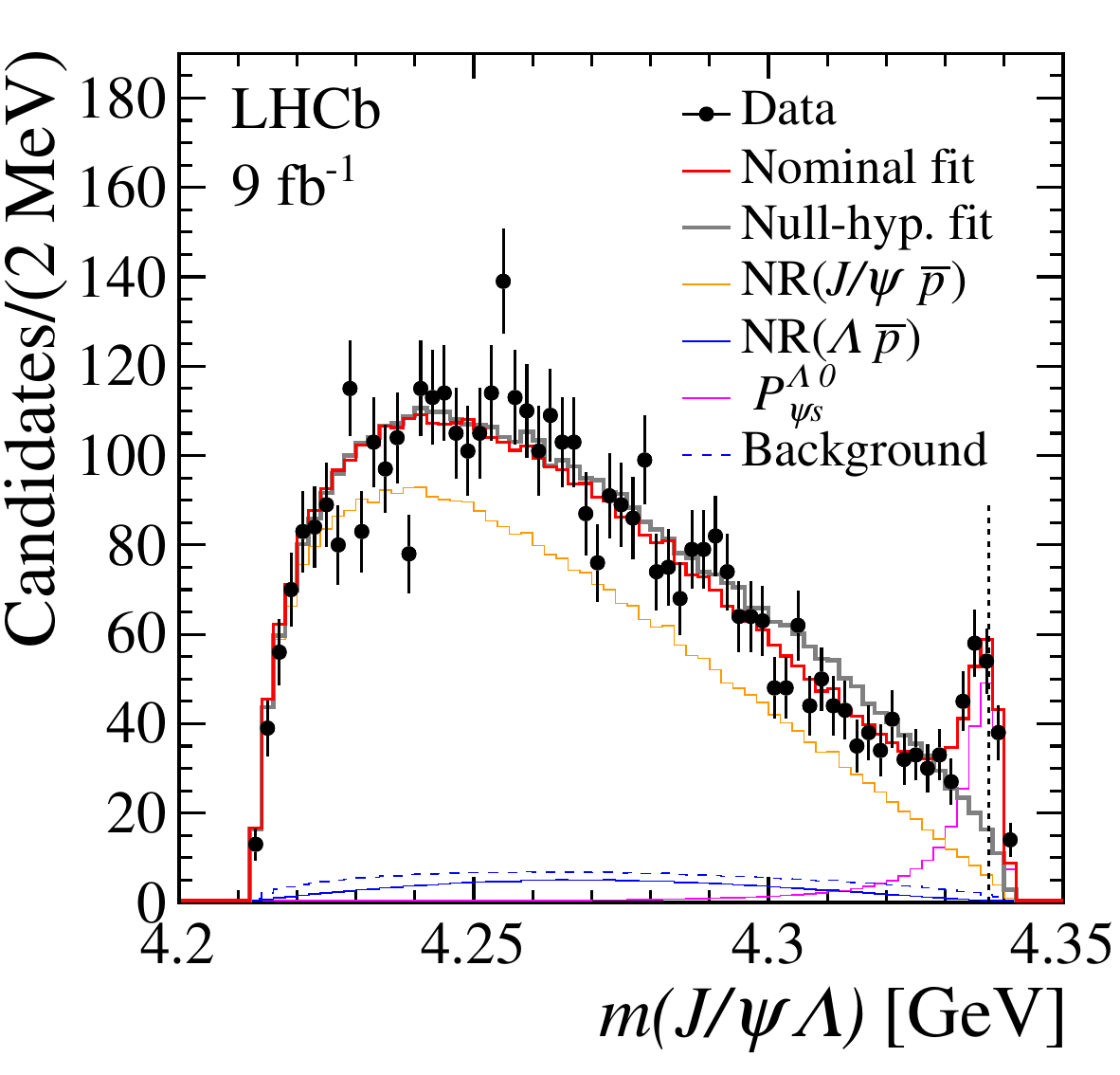}
\ifthenelse{\boolean{captionmode}}{
\wrapcaptionoffset
\caption{Discovery of the $P_{\c\cbar \s}(4338)^0$ at LHCb~\cite{LHCb-PAPER-2022-031}.}
}{}
\end{wrapfigure}

\cardproperty{states}{
\cardnamelist{
 \item $I(J^P)=0(\nicefrac{1}{2}^-)$:  $P_{\c\cbar \s}(4338)^0$
 \item $I(J^P)=0(?)$: $P_{\c\cbar \s}(4458)^0$
 }
}

\cardproperty{minimal quark content}{$[\c\cbar\u\d\s]$}

\cardproperty{experiments}{LHCb}

\cardproperty{production}{$\Bm\to\jpsi\Lambda\bar{\proton}$ ($P_{\c\cbar \s}(4338)$), \\ $\Xi_b\to\jpsi\Lambda\Km$ ($P_{\c\cbar \s}(4458)$)
}

\cardproperty{decay modes}{$\jpsi \Lambda$}

\cardproperty{nearby thresholds}{$\Xi_c^+ \Dm$, $\Xi_c^0 \bar{\D}^{\ast 0}$}

\cardproperty{characteristic widths}{7-17 \mev}

} 

\ifthenelse{\boolean{formpla}}{
\vspace{4mm}
}{
\vspace{12mm}
}

\ifthenelse{\boolean{captionmode}}{\vspace{3mm}}{}

\cardnotes{
In the presence of non-strange hidden-charm pentaquark states, pentaquark states with strangeness are expected due to $SU(3)$ flavour symmetry~\cite{Wu:2010jy,Santopinto:2016pkp,Chen:2016ryt,Shen:2019evi,Xiao:2019gjd,Wang:2019nvm}. The first observation of such a pentaquark candidate, the $P_{\c\cbar \s}(4338)$, with strange quark content was made by LHCb in $B^-\to\jpsi\Lambda \bar{p}$ decay~\cite{LHCb-PAPER-2022-031} which, in fact, has been suggested as a hunting-ground for exotic hadrons long ago~\cite{Brodsky:1997yr}. With a full amplitude analysis, a narrow resonance is observed in the $\jpsi\Lambda$ system at a mass of $4338.2\mev$ and with a fairly small width of ${7.0}\mev$. The spin is determined to be $1/2$ and negative parity is preferred over the positive one at 90\% confidence level. The resonance is observed just a few \mev below the $\Xi_c^+ \Dm$ threshold. The closeness to that threshold as well as the narrow width and the $1/2^-$ spin-parity assignment support a molecular interpretation~\cite{Karliner:2022erb}, but also prompting interpretation as triangle singularity~\cite{Burns:2022uha}.

Another candidate for a pentaquark state in the $\jpsi\Lambda$ system, the $P_{\c\cbar \s}(4458)$, is seen in $\Xi_b\to\jpsi\Lambda\Km$ with a significance just above 3$\sigma$~\cite{Pcs_2021}. 
The enhancement sits $19\mev$ below the $\Xi_c^0 \bar{\D}^{\ast 0}$ threshold and is consistent with the hypothesis of a two-peak structure with spin-parities of $J^P=1/2^-$ and $3/2^-$, as is predicted by molecular models~\cite{Chen:2016ryt,Shen:2019evi,Xiao:2019gjd,Wang:2019nvm,Karliner:2022erb} in analogy to the two-peak structure of $P_{\c\cbar}(4440)$ and $P_{\c\cbar}(4457)$ seen in the non-strange sector below the $\Sigma_c^+ \bar{D}^{*0}$ threshold. Following the same analogy, the $P_{\c\cbar \s}(4338)$ state could be the $SU(3)$ flavor-partner of the $P_{\c\cbar}(4312)$ state. However, due to low statistics it is at present not possible to be conclusive about the two-peak structure. 
}

%% file: conclusion.tex
\section{Conclusion}

The discovery of the $\theX$ in 2003 revived the interest in hadron spectroscopy, showing that even a system as simple as charmonium is far from being understood. 
Today, the field of exotic hadrons is very active,
with many of the candidates only being discovered very recently and in many cases awaiting confirmation. 
The sheer amount of new exotic phenomena that are still being discovered whenever a new region of phase space, especially close to two-body thresholds, becomes available to experiment clearly indicates that we are still exploring which combinations of quarks and gluons are realized as hadrons.
It should be noted that not every new bump seen in experimental data is necessarily a new exotic resonance. On rare occasions, such a claim based on a single observation may be possible, but in most cases independent confirmation from another experiment, in another production mechanism, or using a different decay mode is needed.

While new discoveries are exciting, reaching a full understanding is much more challenging due to the non-perturbative nature of the strong interaction at hadronic mass scales. At present, multiple interpretations -- molecules, compact multiquarks, hybrids, kinematic effects, or simply conventional hadrons -- are considered possible for almost any exotic hadron candidate with little established consensus in the community. 
It is therefore important to determine those observables that most decisively distinguish between the different configuration hypotheses or quantify the relative role of each configuration. 

With the ongoing program of the \BelleTwo, BESIII, GlueX, LHCb and other LHC experiments and upcoming experiments at GSI, the exploration will continue, with more surprises certain to come.
In turn, higher operational luminosities planned for \BelleTwo, BESIII and LHCb in the future will allow us to enter an era of precision measurements involving well-known exotic hadrons in order to advance our understanding of their structure. 